\documentclass[twocolumn,aps,prd,showpacs,showkeys,nofootinbib]{revtex4}
\usepackage{amsmath}
\usepackage{amssymb}
\usepackage{graphicx}
\usepackage{hyperref}
\usepackage{color}
\hypersetup{colorlinks=true,citecolor=blue, linkcolor=red, urlcolor=magenta, hidelinks}

\begin{document}

	\title{Constraining the curvature density parameter in cosmology}
	
	\author{Purba Mukherjee$^1$} 
	\email{pm14ip011@iiserkol.ac.in} 
	
	\author{Narayan Banerjee$^2$}
	\email{narayan@iiserkol.ac.in}

	\affiliation{\\~\\$^1$$^{,2}$Department of Physical Sciences,~~\\Indian Institute of Science Education and Research Kolkata,\\ Mohanpur, West Bengal - 741246, India.}

\begin{abstract}
The cosmic curvature density parameter has been constrained in the present work independent of any background cosmological model. 
The reconstruction is performed adopting the non-parametric Gaussian Processes (GP). The constraints on $\Omega_{k0}$ are 
obtained via a Markov Chain Monte Carlo (MCMC) analysis. Late-time cosmological probes viz., the Supernova (SN) distance modulus  
data, the Cosmic Chronometer (CC) and the radial Baryon Acoustic Oscillations ($r$BAO) measurements of the Hubble data have been 
utilized for this purpose. The results are further combined with the data from redshift space distortions (RSD) which studies the 
growth of large scale structure in the universe. The only \textit{a priori} assumption is that the universe is homogeneous and 
isotropic, described by the FLRW metric. Results indicate that a spatially flat universe is well consistent in 2$\sigma$ within 
the domain of reconstruction $0<z<2$ for the background data. On combining the RSD data we find that the results obtained are 
consistent with spatial flatness mostly within 2$\sigma$ and always within 3$\sigma$ in the domain of reconstruction $0<z<2$.
\end{abstract}

\vskip 1.0cm



\maketitle

\section{Introduction}

The universe on a large scale is described by the spatially homogeneous and isotropic Friedmann-Lema\^{i}tre-Robertson-Walker (FLRW) metric, 

\begin{equation}\label{flrw}
\small ds^2 = - c^2 d t^2 + a^2(t) \left[\frac{d r^2}{1-kr^2} + r^2 d\theta^2 + r^2 \sin^2\theta d\phi^2 \right].
\end{equation}

The scale factor $a(t)$ is the only unknown function to be determined by the field equations. The isotropy and homogeneity of the space section 
demand the spatial curvature to be a constant, which can thus be scaled to pick up values from $-1, +1, 0$. This constant spatial curvature 
is termed the curvature index and is denoted as $k$. This index is not determined by the field equations but is rather fixed by hand, essentially 
from observational requirements. \\

The effect of the spatial curvature $k$ in the evolution of the universe is estimated through the curvature density parameter, defined as, 
\begin{equation}
\Omega_{k} = -\frac{k c^2}{a^2 H^2},
\end{equation}

where $H = \frac{\dot {a}}{a}$ is the Hubble parameter.  $\Omega_{k}$ is positive, negative or zero corresponding to $k = -1, +1, 0$, which in turn correspond 
to open, closed and flat space sections respectively.\\

For the standard cosmological model to correctly describe the present state of the evolution, the initial value of $\Omega_k$ has to be tantalizingly close 
to zero, indicating that the universe essentially starts with a zero spatial curvature. This is known as the flatness or fine-tuning problem for the standard 
cosmology which is believed to be taken care of by an early accelerated expansion called inflation.  For a brief but systematic description, we refer to the 
monograph by Liddle and Lyth\cite{liddle}. Indeed inflation can wash out an early effect of spatial curvature, in comparison with the inflaton energy and the 
Hubble expansion. However, if $\Omega_k$ is negligible but $k$ itself is non-zero, it may reappear in course of evolution and make its presence felt as the 
universe evolves. Reconstruction of some dark energy parameters indicate that a non-flat space section may not be easily ruled out. The use of $\Omega_{k}$ 
as a free parameter is found to affect the reconstruction of dark energy equation of state parameter $w(z)$, as shown by Clarkson, Cortes, and 
Bassett\cite{clarkson2007}. A reconstruction of the deceleration parameter $q(z)$ by Gong and Wang\cite{gong2007} shows that although a flat universe is still 
consistent, $|\Omega_{k0}|$ is less than only 0.05 for a one-parameter dark energy model and lies between -0.064 and 0.028 for a $\Lambda$CDM model with spatial 
curvature, where a subscript 0 indicates the present value of the quantity. The recent Planck\cite{planck} data also indicates that a universe with a non-zero 
spatial curvature may not be completely ruled out.  \\

The motivation of the present work is to constrain the curvature density parameter $\Omega_{k0}$ hence attempt to ascertain the signature of the curvature index 
$k$, directly from observational data without assuming any background cosmological model. We do not start from any theory of gravity or use any form of matter 
distribution in the universe. The only \textit{a priori} assumption is that the universe is homogeneous and isotropic, and thus described by the FLRW metric. 
There is quite a lot of interest in this direction, which is normally pursued along with constraining other cosmological parameters pertaining to the alleged 
accelerated expansion of the universe. Most of these investigations depend on some chosen parametric form of cosmological quantities related to the late-time 
expansion behaviour of the universe\cite{leo2016,witz2018,deni2018,cao2019,li2020,gratton2020,park2020,nunes2020,benisty2021,handly2021}. This approach is 
indeed biased by the parametrization, as the functional form of the quantity is already chosen. \\

Another way of reconstruction involves a verification of the FLRW metric from datasets, and ascertaining the value of $\Omega_{k0}$ as a by-product by 
combining the dimensionless reduced Hubble parameter $E(z)$ and the normalised comoving distance $D(z)$\cite{clarkson2007,clarkson2008,cai2016,rana2017,
liu2020,arjona2021,ref_new2014}. \\

The present work does not assume any functional form of $\Omega_k$, but rather resorts to a non-parametric reconstruction of $\Omega_{k0}$, the present 
value of the curvature density parameter. The idea is to obtain constraints on the geometrical quantity $\Omega_{k0}$ using recent observational data provided by 
the high precision cosmological probes, namely, the Supernova (SN) distance modulus data, the Cosmic Chronometer (CC) and the radial Baryon Acoustic Oscillations 
($r$BAO) measurements of the Hubble parameter. We also combine these data from background measurements with the data from redshift space distortions (RSD) due to 
the growth of large scale structures. The reconstruction is performed adopting the non-parametric Gaussian Processes (GP). The resulting marginalized 
constraints on $\Omega_{k0}$ are obtained via a Markov Chain Monte Carlo (MCMC) analysis, independent of any parametric model of the expansion history. \\

Attempts towards obtaining constraints on $\Omega_{k0}$ using the non-parametric approach started to gain momentum in the recent past. 
Li \textit{et al.}\cite{li2016}, Wei and Wu\cite{wei2017} proposed to constrain the cosmic curvature in a model-independent way by combining the CC-$H(z)$ with 
Union 2.1\cite{union2.1}, and Joint Light-curve Analysis (JLA)\cite{jla} SN-Ia data respectively. Model-independent constraints on cosmic curvature and opacity 
was carried out by Wang \textit{et. al.}\cite{wang2017} using the CC-H(z) and JLA SN-Ia data. Liao\cite{liao2019} studied constraints on cosmic curvature with 
lensing time delays and gravitational waves (GWs). Model-independent distance calibration and $\Omega_{k0}$ measurement using Quasi-Stellar Objects (QSOs) and 
CCs was done by Wei and Melia\cite{wei2019}. Ruan \textit{et al.}\cite{ruan2019} obtained constraints on $\Omega_{k0}$ using the CC-$H(z)$ data and HII galaxy 
Hubble diagram. Model-independent estimation for $\Omega_{k0}$ from the latest strong gravitational lens systems (SGLs) was performed by Zhou and Li\cite{zhou2019}. 
Wang \textit{et al.}\cite{wang2019} constrained $\Omega_{k0}$ from SGL and Pantheon\cite{pan1} SN-Ia observations. Wang, Ma and Xia\cite{wang2020} employed a machine 
learning algorithm called Artificial Neural Network (ANN) to constrain $\Omega_{k0}$ using data from CC, SN-Ia and GWs. Recently, Yang and Gong\cite{yang2021} 
constrained the $\Omega_{k0}h^2$ using CC-$H(z)$, Pantheon SN-Ia and RSD data where $h = \frac{H_0}{100~ \mbox{\small km}~\mbox{\small Mpc}^{-1} ~\mbox{\small 
		s}^{-1}}$ is the dimensionless Hubble parameter at the present epoch. Non-parametric spatial curvature inference using CC and Pantheon data was performed 
by Dhawan, Alsing and Vagnozzi\cite{dhawan2021}. A majority of these investigations use GP as their numerical tool.\\

We use observational data more recent than most of these investigations, but the major difference is that we include a wider variety of data in combination, 
measuring different features of the evolution. We also include a section where the RSD dataset which has mostly eluded the attention so far, except the work 
of Yang and Gong\cite{yang2021} in the reconstruction of $\Omega_{k0}h^2$ despite its utmost relevance in this connection, as the growth of perturbations has 
to be consistent with the spatial curvature. \\

The other crucial addition in the present work is that we also check the consistency of the constraints on spatial curvature with thermodynamic requirements. 
Very recently, Ferreira and Pav\'{o}n\cite{pavon} imposed a relation using the generalized second law of thermodynamics, which reads as $1 + q \geq \Omega_{k}$, 
where $q$ is the deceleration parameter. It is quite reassuring to see that constraints on $\Omega_{k0}$ quite comfortably satisfies the requirement. \\

The results obtained indicate that a spatial curvature may indeed exist at the present epoch. But the estimated sign of the curvature depends on the strategies 
for measuring $H_0$ to some extent. But the results are statistically not too significant, as a zero curvature is mostly included in 1$\sigma$ and always at 
least in 2$\sigma$.\\

The paper is organized as follows. Section 2 contains the details on the reconstruction method. In section 3, the observational data used in the present work 
have been briefly reviewed. The methodology is discussed in section 4. Reconstruction using background data is performed in section 5. Section 6 shows the 
consistency of $\Omega_{k0}$ constraints with the second law of thermodynamics. Reconstruction using the perturbation data are presented in section 7. The final 
section 8 contains an overall discussion on the results. 

\section{Gaussian Process}

We shall employ the well-known Gaussian processes (GP)\cite{william, mackay, rw} for the reconstruction of ${\Omega}_{k0}$. Assuming that the observational 
data obey a Gaussian distribution with mean and variance, the posterior distribution of the reconstructed function (say $f$) and its derivatives can be 
expressed as a joint Gaussian distribution. In this method, the covariance function $\kappa(z, \tilde{z})$ plays a key role. It correlates the values of 
$f(z)$ at two redshift points $z$ and $\tilde{z}$. This covariance function depends on a set of \textit{hyperparameters} which are optimised by maximizing 
the log marginal likelihood. With the optimised covariance function, the data can be extended to any redshift point. The GP method has been widely applied 
in cosmology \cite{gp0,gp1,gp2,gp3,gp4,gp5,gp6,gp7,gp8,gp9,gp10,keeley2020,xia_q,lin_q,jesus_q,purba_q,purba_cddr,purba_j,purba_int,benisty,kamal}.\\

It deserves mention that the choice of $\kappa(z, \tilde{z})$, affects the reconstruction to some extent. The more commonly used covariance function is the 
squared exponential covariance,  which is infinitely differentiable,

\begin{equation} 
\kappa(z, \tilde{z}) = \sigma_f^2 \exp \left( - \frac{(z-\tilde{z})^2}{2l^2}\right). \label{sqexp}
\end{equation} 

In this particular work we consider the squared exponential, Mat\'{e}rn 9/2, Cauchy and rational quadratic covariance functions. The Mat\'{e}rn 9/2 covariance 
function is given by,

\begin{eqnarray} 
\kappa(z,\tilde{z}) = \sigma_f^2 \exp \left( \frac{-3 \vert z - \tilde{z} \vert}{l} \right) \left[ 1 + \frac{3 \vert z - \tilde{z} \vert}{l} + \right. \nonumber \\
\left. + \frac{27 \left( z - \tilde{z} \right)^2}{7l^2} + \frac{18 \vert z - \tilde{z} \vert ^3}{7l^3} + \frac{27 \left( z - \tilde{z} \right)^4}{35l^4}\right] . \label{mat92}
\end{eqnarray}

The Cauchy covariance function is

\begin{eqnarray}
\kappa(z,\tilde{z}) = \sigma_f^2 \left[ \frac{l}{(z-\tilde{z})^2 + l^2}\right] ,  \label{cauchy}
\end{eqnarray}

and the rational quadratic covariance function is

\begin{eqnarray}
\kappa(z,\tilde{z}) = \sigma_f^2 \left[ 1 + \frac{(z-\tilde{z})^2}{2 \alpha l^2} \right]^{-\alpha} ,  \label{ratquad}
\end{eqnarray} 

where $\sigma_f$ , $l$ and $\alpha$ are the kernel hyperparameters. Throughout this work, we assume a zero mean function \textit{a priori} to characterize the GP. \\

For more details on the GP method, one can refer to the Gaussian Process website\footnote{\url{http://www.gaussianprocess.org}}. The publicly 
available  \texttt{GaPP}\footnote{\url{https://github.com/carlosandrepaes/GaPP}} (Gaussian Processes in python) code by Seikel \textit{et al.}\cite{gp1} has been 
used in this work.

\section{Observational Data}

In this work we use both the background data and the perturbation data for the reconstruction of ${\Omega}_{k0}$. The background level includes different 
combinations of datasets involving the Cosmic Chronometer data (CC), the Supernova distance modulus data (SN), the Baryon Acoustic Oscillation data (BAO). 
For the perturbation level data, the growth rate of structure $f \sigma_8$ from the redshift-space distortions (RSD) are utilized. A brief summary of 
the datasets is given below.

\subsection{Background Level}

The Hubble parameter $H(z)$ can be directly obtained from the differential redshift time derived by calculating the spectroscopic differential ages of 
passively evolving galaxies, usually called the Cosmic Chronometer (CC) method \cite{jimenez2002}. In this work we use the latest 31 CC $H(z)$ data 
\cite{cc0,cc1,cc2,cc3,cc4,cc5,cc6}, covering the redshift range up to $z \sim 2$. These measurements do not assume on any particular cosmological 
model. \\

We take into account the updated and corrected Pantheon compilation by Steinhardt, Sneppen and Sen\cite{pan_correct}. This corrected sample improves 
upon some errors in the quoted values of the redshift $z$ in the original Pantheon dataset by Scolnic \textit{et al.}\cite{pan1}. The Pantheon 
catalogue is presently the largest spectroscopically confirmed SNIa sample, consisting of 1048 supernovae from different surveys covering the redshift 
range up to $z \sim 2.3$, including the SDSS, SNLS, various low-$z$ and some high-$z$ samples from the HST. \\

An alternative compilation of the Hubble $H(z)$ data can be deduced from the radial BAO peaks in the galaxy power spectrum, or from the BAO peak using 
the Ly-$\alpha$ forest of quasars, which are based on the clustering of galaxies or quasi stellar objects (namely $r$BAO), spanning the redshift range 
$ 0 < z < 2.4$ reported in various surveys \cite{bao0,bao1,bao2,bao3,bao4,bao5,bao6,bao7,bao8,bao9,bao10,bao11,bao12}. One may find that some of the 
$H(z)$ data points from clustering measurements are correlated since they either belong to the same analysis or there is an overlap between the galaxy 
samples. Here in this paper, we mainly consider the central value and standard deviation of the data into consideration. Therefore, we assume that they 
are independent measurements as in \cite{rsd_comp, purba_j}.\\

In view of the known tussle between the value of $H_0$ as given by the Planck\cite{planck} 2018 data from the CMB measurements (hereafter referred to as 
P18), and that from HST observations of 70 long-period Cepheids in the Large Magellanic Clouds by the SH0ES\cite{riess1} team (hereafter referred to as R19), 
reconstruction using both of them have been carried out separately. The recent global P18 and local R19 measurements of $H_0 = 67.27 \pm 0.60 $ km s$^{-1}$ 
Mpc$^{-1}$ for TT+TE+EE+lowE (P18)\cite{planck} and $H_0 = 74.03 \pm 1.42 $ km s$^{-1}$ Mpc$^{-1}$ (R19)\cite{riess1} respectively, with a $4.4 \sigma$ 
tension between them, are considered for the purpose. 

\subsection{Perturbation Level}

The redshift space distortion (RSD) data is a very promising probe to distinguish between different cosmological models. Various dark energy models 
may lead to a similar evolution in the large scale but can show a distinguishable growth of the cosmic structure. In this work, we utilize the updated 
datasets of the $f \sigma_8$ measurements, including the collected data from 2006-2018 \cite{rsd0,rsd1,rsd2,rsd3,rsd4}, and the completed SDSS, 
extended BOSS Survey, DES and other galaxy surveys \cite{rsd5,rsd6,rsd7,rsd8,rsd9,rsd10,rsd11,rsd12,rsd13,rsd14,rsd15,rsd16,rsd17,rsd18,rsd19,rsd20,
	rsd21,rsd22,rsd23,rsd24,rsd25}. We refer to \cite{rsd_comp} for a recent compilation of the 63 RSD data within the redshift range $0<z<2$ respectively. 
This $f \sigma_8$ is called the growth rate of structure.

\section{The curvature density parameter and distance measures}

In an FLRW universe, the proper distance from the observer to a celestial object at redshift $z$ along the line of sight is given by,

\begin{equation}\label{dp}
d_p (z)= \frac{c}{H_0} \int_{0}^{z} \frac{d z'}{E(z')}
\end{equation} 

and the transverse comoving distance can be expressed as,

\begin{equation} \label{dM}
d_M(z)= \frac{c}{H_0\sqrt{\vert \Omega_{k0} \vert}}\sin\mbox{$n$} \left( \sqrt{\vert \Omega_{k0} \vert } \int_{0}^{z} \frac{d z'}{E(z')}\right) ,
\end{equation}

in which the $\sin n$ function is a shorthand for,

\[ \sin nx= \begin{cases}
\sinh x &  (\Omega_{k0}>0), \\
~~x & (\Omega_{k0} \rightarrow 0), \\
\sin x & (\Omega_{k0}<0).
\end{cases}
\]

We define the reduced Hubble parameter as,

\begin{equation}\label{h_recon}
E(z) = \frac{H (z)}{H_0}.
\end{equation} 

Here, a suffix 0 indicates the value of the relevant quantity at the present epoch and $z$ is the redshift, defined as $1+z \equiv \frac{a}{a_0}$. 
The dimensionless parameter $\Omega_{k0}$, namely the cosmic curvature density parameter, defined as

\begin{equation}\label{omeggak-def}
\Omega_{k0} = -\frac{k c^2}{a_0^2 H_0^2},
\end{equation}

is positive, negative or zero corresponding to the spatial curvature $k = -1, +1, 0$ which signifies an open, closed, or flat universe, 
respectively.\\

For convenience, we can define the normalized proper distance,

\begin{equation}\label{Dp}
D_p(z) \equiv \frac{ H_0 }{c}~d_p(z)
\end{equation} 

and the normalized transverse comoving distance,

\begin{equation}\label{D}
D(z) \equiv \frac{ H_0 }{c}~d_M(z)
\end{equation} 

as dimensionless cosmological distance measures which will be used later in our work.

\section{Reconstruction from Background data}

In the very beginning we use the GP method to reconstruct the Hubble parameter $H(z)$ from the CC data and CC+$r$BAO data. We then normalize the datasets 
with the reconstructed value of $H_0$ i.e., $H(z=0)$ to obtain the dimensionless or reduced Hubble parameter $E(z)$. Considering the error associated with 
the Hubble data as $\sigma_{H}$, we calculate the uncertainty in $E(z)$ as,

\begin{equation} \label{sigh_recon}
{\sigma_{E}} = \sqrt{\frac{{\sigma_H}^2}{ {H_0}^2} + \frac{H^2}{{H_0}^4}{\sigma_{H_0}}^2},
\end{equation} 

where $\sigma_{H_0}$ is the error associated with $H_0$. \\

With the function $E(z)$ reconstructed from the Hubble data, as described in equation \eqref{h_recon}, the normalised proper distance $D_p$ is calculated via 
a numerical integration using the composite trapezoidal\cite{trapez} rule.

\begin{eqnarray} \label{trap}
D_p(z) &=& \int_{0}^{z} \frac{d z'}{E(z')} \nonumber \\ 
&\simeq& \frac{1}{2} \sum_{i} (z_{i+1} - z_i) \left[\frac{1}{E(z_{i+1})}+\frac{1}{E(z_{i})}\right].
\end{eqnarray} 

Thus we get $D_p$ without assuming any prior fiducial cosmological model. The error associated with $D_p$, say $\sigma_{D_p}$, is obtained from the reconstructed 
function $E(z)$ along with its associated error uncertainties $\sigma_{E}(z)$ described in equation \eqref{sigh_recon}, and is given by,  

\begin{equation}
\sigma^2_{D_p}(z) \simeq \frac{1}{4} \sum_{i} (z_{i+1} - z_i)^2 \left[\frac{\sigma^2_{E_{i+1}}}{E^4_{i+1}}+\frac{\sigma^2_{E_{i}}}{E^4_{i}}\right].
\end{equation}

From this reconstructed $D_p$, we can calculate the normalised transverse comoving distance $D$ from the Hubble data as,

\begin{eqnarray} \label{D_recon}
D(z)= \begin{cases}
\frac{1}{\sqrt{\Omega_{k0}}}\sinh \left[ \sqrt{\Omega_{k0}} ~ D_p(z)\right] & \Omega_{k0}>0, \\
~~D_p(z) & \Omega_{k0}=0, \\
\frac{1}{\sqrt{-\Omega_{k0}}}\sin \left[ \sqrt{-\Omega_{k0}} ~ D_p(z)\right] &  \Omega_{k0}<0. 
\end{cases}
\end{eqnarray} 

The error $\sigma_D$ of the reconstructed $D$ from the Hubble data is,

\begin{eqnarray} \label{sigD_recon}
\sigma_{D}(z)= \begin{cases}
\cosh \left[ \sqrt{\Omega_{k0}} ~ D_p(z)\right] \sigma_{D_p}(z) & \Omega_{k0}>0, \\
~~\sigma_{D_p}(z) & \Omega_{k0}=0, \\
\cos \left[ \sqrt{-\Omega_{k0}} ~ D_p(z)\right] \sigma_{D_p}(z) &  \Omega_{k0}<0.
\end{cases}
\end{eqnarray}

Steinhardt \textit{et al.}\cite{pan_correct} lists the corrected distance modulus $\mu$ corresponding to different redshift $z$, along with their 
respective error uncertainties, from supernovae observations following the BEAMS with Bias Corrections (BBC)\cite{beams} framework. \\

The total uncertainty matrix of observed distance modulus given by, 
\begin{equation}
\mathbf{\Sigma}_{\mu} = \mathbf{C}_{\mbox{\tiny stat}} + \mathbf{C}_{\mbox{\tiny sys}} ,
\end{equation} 

where both the statistical covariance matrix $\mathbf{C}_{\mbox{\tiny stat}}$ and the systematic errors $\mathbf{C}_{\mbox{\tiny sys}}$ are included 
in our calculation.\\

With another Gaussian Process on the observed distance modulus of the SN-Ia data, we reconstruct $\mu_{\mbox{\tiny SN}}$ and the associated error 
uncertainties ${\sigma}_{\mu_{\mbox{\tiny SN}}}$, at the same redshift as that of the Hubble data. The subscript {\small SN} stands for supernova.\\

The distance modulus is theoretically given by,

\begin{equation}  \label{mu}
\mu = 5 \log_{10} \left(\frac{d_L}{\mbox{Mpc}}\right) + 25.
\end{equation} 

Here, $d_L$ is the luminosity distance. This $d_L$ is related to the normalised transverse comoving distance $D$ as,

\begin{equation}
d_L(z) = d_M (1+z) = \frac{c(1+z) D}{H_0} .
\end{equation}

Substituting equation \eqref{D_recon} in equation \eqref{mu}, we estimate the reconstructed distance modulus from the Hubble data, say $\mu_{\mbox{\tiny H}}$ 
along with its 1$\sigma$ error uncertainty $\sigma_{\mu_{\mbox{\tiny H}}}$ as,

\begin{eqnarray}
\mu_{{\mbox{\tiny H}}} &=& 5 \log_{10} \left[ \frac{c (1+z) D}{H_0}\right] + 25,  \label{mu_H}\\
\sigma_{\mu_{\mbox{\tiny H}}} &=& \frac{5}{\ln 10} \frac{\sigma_{D}}{D} \label{sigmu_H}.
\end{eqnarray}

Equations \eqref{D_recon}, \eqref{sigD_recon}, \eqref{mu_H} and \eqref{sigmu_H} will finally be utilized for obtaining the contour plots between $\Omega_{k0}$ 
and $H_0$ at different confidence levels.\\

Finally we constrain the curvature density parameter $\Omega_{k0}$ and the Hubble parameter $H_0$ simultaneously by minimizing the 
$\chi^2$ statistics. The $\chi^2$ function is given by, 

\begin{equation}
\chi^2 = \Delta \mu^T ~ \mathbf{\Sigma}^{-1} ~ \Delta \mu .
\end{equation} 

$\Delta \mu = \mu_{{\mbox{\tiny SN}}} - \mu_{{\mbox{\tiny H}}}$ is the difference between the distance moduli of Pantheon SN-Ia and that of the $H(z)$ data. 
$\mathbf{\Sigma} = {\sigma}^2_{\mu_{\mbox{\tiny SN}}} + \sigma_{\mu_{\mbox{\tiny H}}}^2 $ is the total uncertainty matrix from combined Pantheon and Hubble 
data.\\ 

We attempt to reconstruct $\Omega_{k0}$ directly for the following combination of data sets,
\begin{itemize}
	\item Set I
	\begin{enumerate}
		\item N1 - CC+SN 
		\item P1 - CC+SN+P18
		\item R1 - CC+SN+R19
	\end{enumerate}	
	\item Set II
	\begin{enumerate}
		\item N2 - CC+$r$BAO+SN 
		\item P2 - CC+$r$BAO+SN+P18
		\item R2 - CC+$r$BAO+SN+R19
	\end{enumerate}	
\end{itemize}

We get the constraints on $\Omega_{k0}$ and $H_0$ along with their respective error uncertainties by a Markov Chain Monte Carlo (MCMC) analysis with the 
assumption of a uniform prior distribution for $\Omega_{k0} \in [-1,1]$ and $H_0 \in [50, 100]$ in case of the N1 and N2 combinations respectively. For 
the P1 and P2 combinations, we consider the P18 Gaussian $H_0$ prior whereas, for R1 and R2 combinations, the R19 Gaussian $H_0$ prior has been used.\\

\begin{figure*} [t!]
\begin{center}
	\includegraphics[angle=0, width=0.285\textwidth]{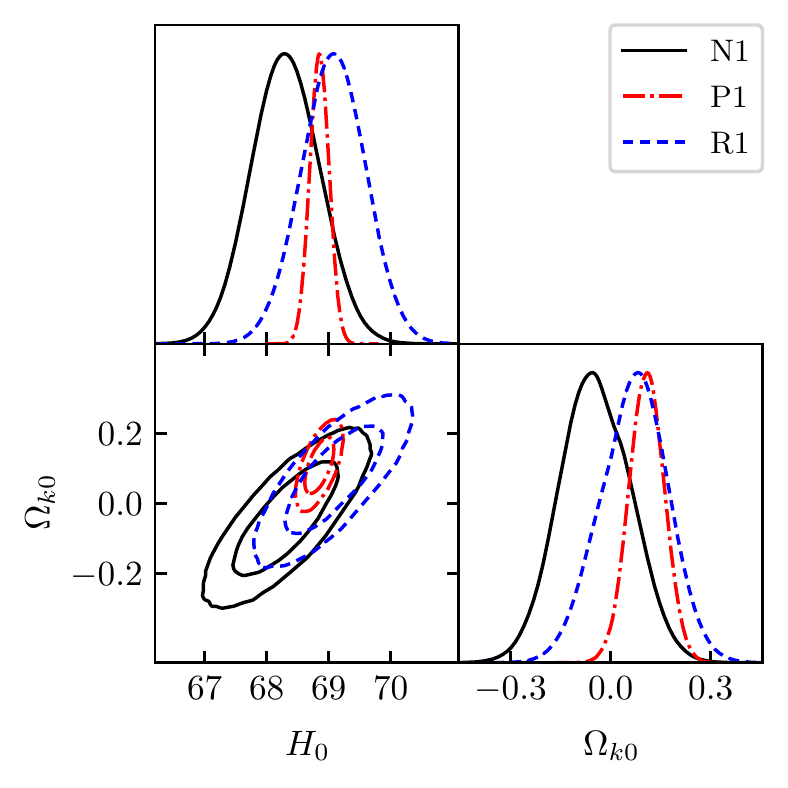} \hspace{2cm}
	\includegraphics[angle=0, width=0.285\textwidth]{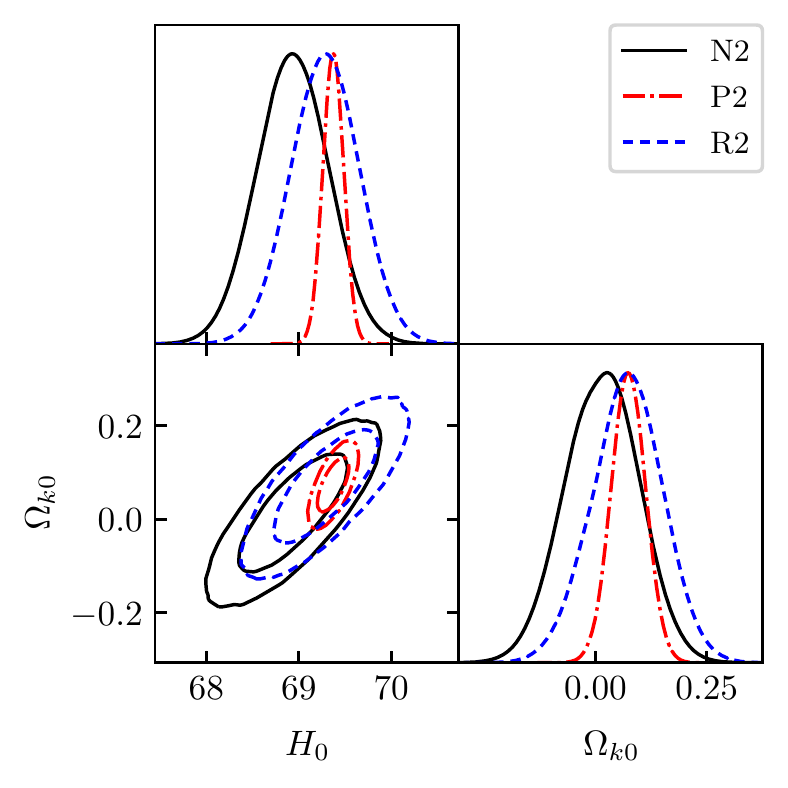} 
\end{center}
	\caption{{\small Contour plots and the marginalized likelihood of $H_0$ and $\Omega_{k0}$ considering the squared exponential covariance for Set I (left) 
			and Set II (right). The solid lines represent the results for N1 and N2 data-set combination, dash-dot lines corresponds to the P1 and P2 data-set 
			combination, and the dashed lines  represent the results for R1 and R2 data-set combinations. The associated 1$\sigma$, 2$\sigma$ confidence contours 
			are shown along with the respective marginalized likelihood functions.}}
	\label{Ok_sqexp}
\end{figure*}

\begin{figure*} [t!]
	\begin{center}
		\includegraphics[angle=0, width=0.285\textwidth]{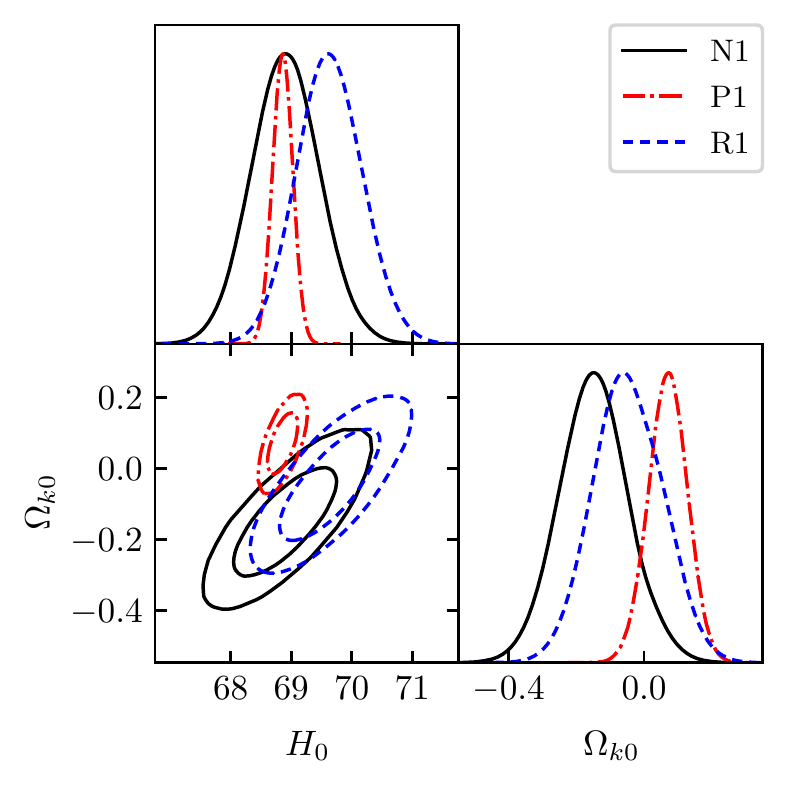} \hspace{2cm}
		\includegraphics[angle=0, width=0.285\textwidth]{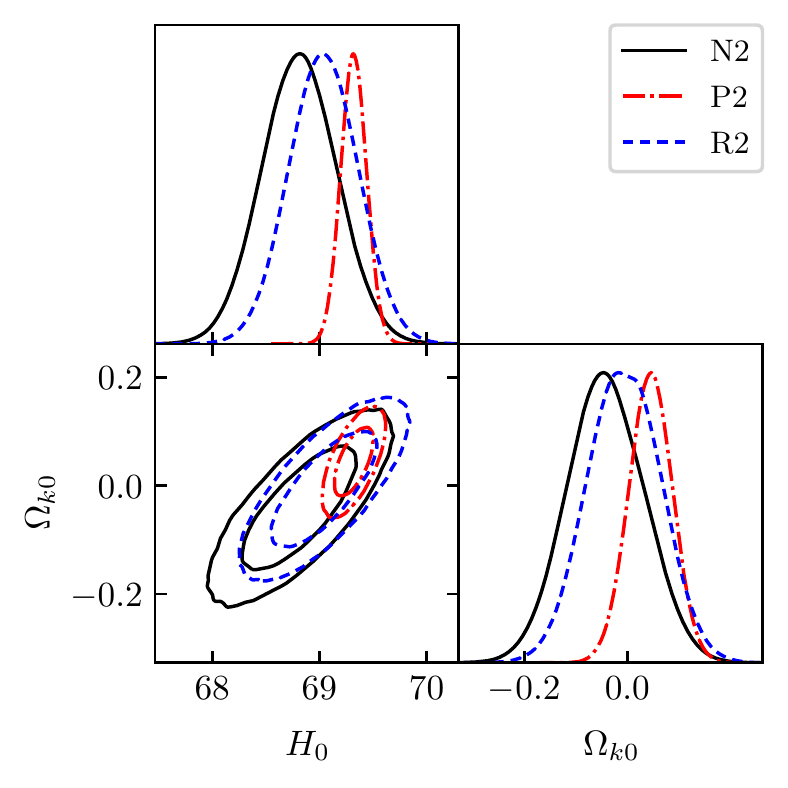}
	\end{center}
	\caption{{\small Contour plots and the marginalized likelihood of $H_0$ and $\Omega_{k0}$ considering the Mat\'{e}rn $9/2$ covariance for Set I (left) 
			and Set II (right). The solid lines represent the results for N1 and N2 data-set combination, dash-dot lines corresponds to the P1 and P2 data-set 
			combination, and the dashed lines  represent the results for R1 and R2 data-set combinations. The associated 1$\sigma$, 2$\sigma$ confidence contours 
			are shown along with the respective marginalized likelihood functions.}}
	\label{Ok_mat92}
\end{figure*}

\begin{figure*} [t!]
	\begin{center}
		\includegraphics[angle=0, width=0.285\textwidth]{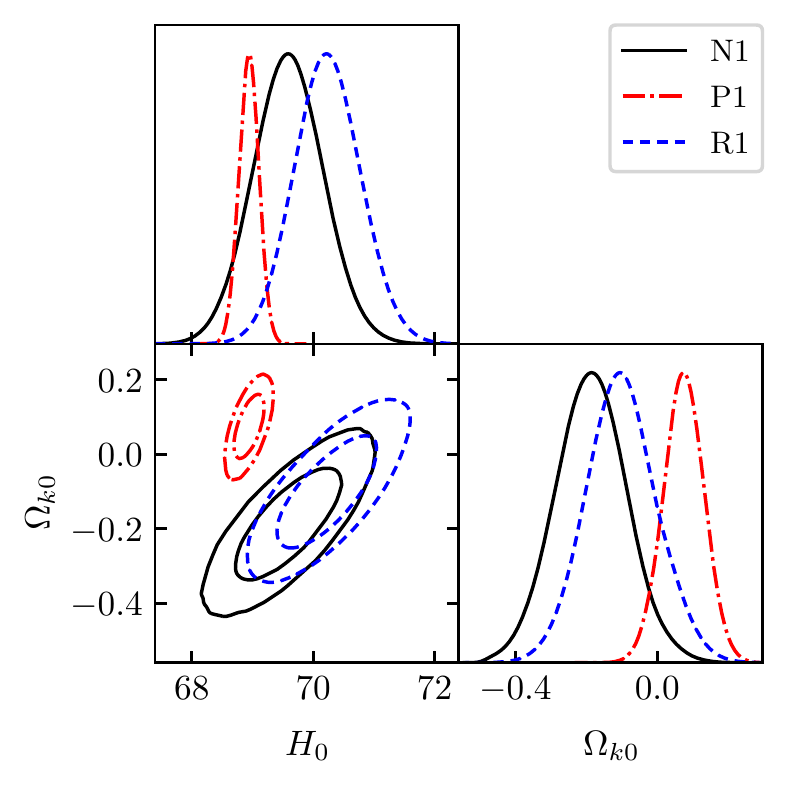} \hspace{2cm}
		\includegraphics[angle=0, width=0.285\textwidth]{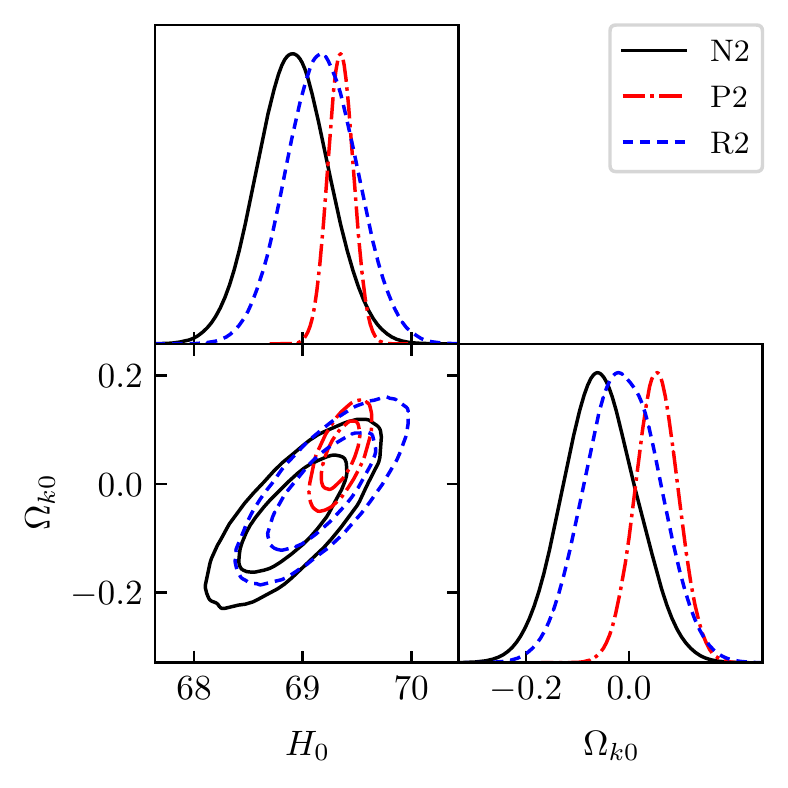} \\
	\end{center}
	\caption{{\small Contour plots and the marginalized likelihood of $H_0$ and $\Omega_{k0}$ considering the Cauchy covariance for Set I (left) 
			and Set II (right). The solid lines represent the results for N1 and N2 data-set combination, dash-dot lines corresponds to the P1 and P2 data-set 
			combination, and the dashed lines  represent the results for R1 and R2 data-set combinations. The associated 1$\sigma$, 2$\sigma$ confidence contours 
			are shown along with the respective marginalized likelihood functions.}}
	\label{Ok_cauchy}
\end{figure*}

\begin{figure*} [t!]
	\begin{center}
		\includegraphics[angle=0, width=0.285\textwidth]{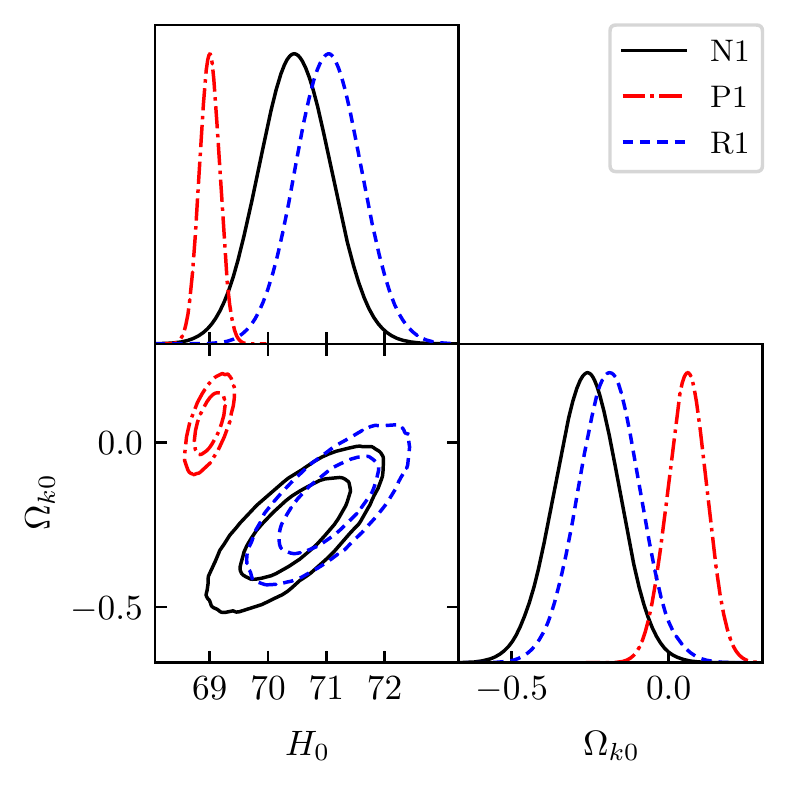} \hspace{2cm}
		\includegraphics[angle=0, width=0.285\textwidth]{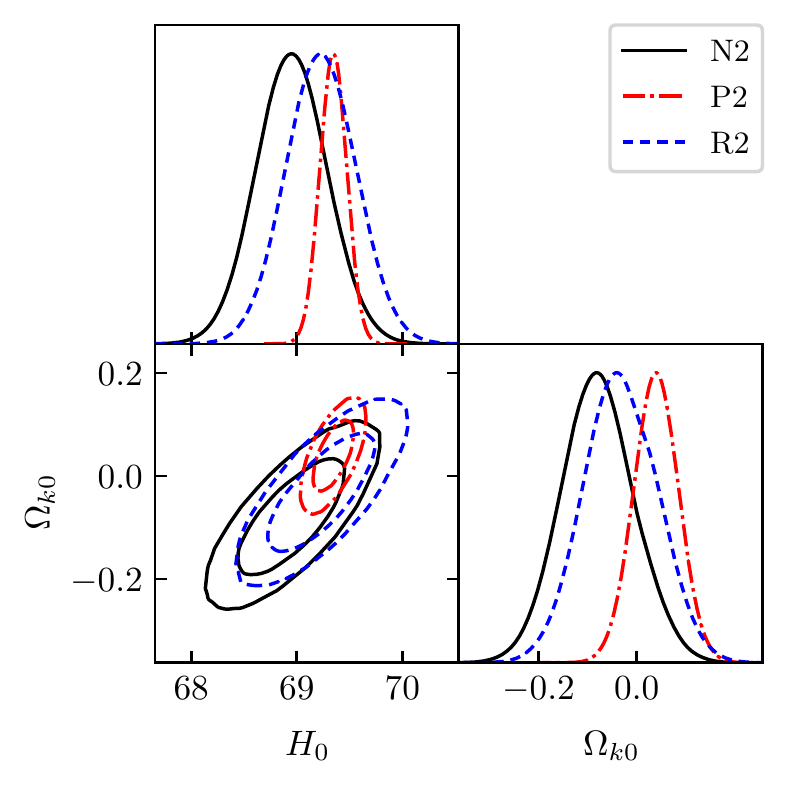}
	\end{center}
	\caption{{\small Contour plots and the marginalized likelihood of $H_0$ and $\Omega_{k0}$ considering the rational quadratic covariance for Set I (left) 
			and Set II (right). The solid lines represent the results for N1 and N2 data-set combination, dash-dot lines corresponds to the P1 and P2 data-set 
			combination, and the dashed lines  represent the results for R1 and R2 data-set combinations. The associated 1$\sigma$, 2$\sigma$ confidence contours 
			are shown along with the respective marginalized likelihood functions.}}
	\label{Ok_quad}
\end{figure*}

In this work, we adopt a python implementation of the ensemble sampler for MCMC, the publicly available \texttt{emcee}\footnote{\url{https://github.com/dfm/emcee}}, 
introduced by Foreman-Mackey \textit{et al.}\cite{emcee}. The best fit results along with their respective 1$\sigma$, 2$\sigma$ and 3$\sigma$ uncertainties is 
given in Table \ref{HOk_res}. We plot the results using the \texttt{GetDist}\footnote{\url{https://github.com/cmbant/getdist}} module of python, developed by 
Lewis\cite{getdist}. The plots for the marginalized distributions with 1$\sigma$ and 2$\sigma$ confidence contours for $\Omega_{k0}$ and $H_0$ are shown in Figures 
\ref{Ok_sqexp}, \ref{Ok_mat92}, \ref{Ok_cauchy} and \ref{Ok_quad} considering the squared exponential, Mat\'{e}rn $9/2$, Cauchy and rational quadratic covariance 
respectively. \\

The reconstructed $\Omega_{k0}$ for the N1 combination are consistent with spatial flatness within 2$\sigma$ confidence level (CL) for the squared exponential, 
Mat\'{e}rn 9/2 and Cauchy covariance functions, and within 3$\sigma$ CL for the rational quadratic covariance. With the addition of $r$BAO data in Set II, the 
constraints on $\Omega_{k0}$ become tighter. From the combined N2 data-set, we find that $\Omega_{k0}$ is consistent with a spatially flat universe at 1$\sigma$ CL 
for the squared exponential, Mat\'{e}rn 9/2 and Cauchy covariance, whereas in 2$\sigma$ for the rational quadratic kernel. The best-fit values shows an inclination 
towards a closed universe for N1 and N2 data-sets. The degeneracy between $H_0$ and $\Omega_{k0}$ along with their correlation has also been shown.\\

We also examine if the two different strategies for determining value of $H_0$, with conflicting results, affect the reconstruction significantly. We plot the 
marginalized distributions with 1$\sigma$ and 2$\sigma$ confidence contours for $\Omega_{k0}$ and $H_0$ using the P1 and P2 combinations considering the P18 
prior on $H_0$, and for the R1 and R2 combinations considering the R19 prior on $H_0$ in figures 1-4. With the inclusion of the P18 data prior we see that the 
best-fit values of $\Omega_{k0}$ favour a spatially open universe, whereas in case of the choice of R19 as a prior, the best-fit values of the constrained 
$\Omega_{k0}$ shows that the combined data favours a spatially closed universe. However, a spatially flat universe is mostly included at 2$\sigma$ CL for both 
cases. \\

\begin{table*}[t!]
	\caption{{\small The best fit values of $\Omega_{k0}$, $H_0$ and reconstructed $1+q_0$ along with their 1$\sigma$, 2$\sigma$ and 3$\sigma$ uncertainties from different combinations of 
			datasets, for four choices of the covariance function using the background data.}}
	\begin{center}
		\resizebox{0.99\textwidth}{!}{\renewcommand{\arraystretch}{1.45} \setlength{\tabcolsep}{28pt} \centering  
			\begin{tabular}{l c c c c } 
				\hline 	
				\textbf{Dataset} & $\kappa(z,\tilde{z})$ &  $H_0$ & $\Omega_{k0}$ & $1+q_0$ \\ 
				\hline 
				\hline
				N1 &  & $68.30^{+0.56~+1.09~+1.67}_{-0.55~-1.06~-1.59}$ & $-0.05^{+0.11~+0.21~+0.32}_{-0.10~-0.20~-0.31}$ & $0.49^{+0.06~+0.12~+0.18}_{-0.06~-0.13~-0.19}$ \\ 
				
				N2 &  & $68.94^{+0.39~+0.76~+1.15}_{-0.39~-0.75~-1.14}$ & $~0.02^{+0.08~+0.16~+0.24}_{-0.09~-0.16~-0.25}$  & $0.42^{+0.03~+0.06~+0.09}_{-0.03~-0.06~-0.09}$\\ 
				
				P1 & Sq. & $68.85^{+0.16~+0.31~+0.46}_{-0.16~-0.31~-0.46}$ & $~0.11^{+0.05~+0.11~+0.16}_{-0.05~-0.11~-0.16}$ & $0.46^{+0.06~+0.12~+0.18}_{-0.06~-0.12~-0.18}$ \\ 
				
				P2 & Exp. &  $69.37^{+0.11~+0.22~+0.33}_{-0.11~-0.22~-0.33}$ & $~ 0.07^{+0.04~+0.08~+0.12}_{-0.04~-0.08~-0.12}$ & $0.39^{+0.03~+0.06~+0.10}_{-0.03~-0.06~-0.10}$ \\ 
				
				R1 &  & $69.08^{+0.52~+1.02~+1.54}_{-0.52~-1.03~-1.55}$ & $~0.07^{+0.10~+0.19~+0.29}_{-0.11~-0.21~-0.31}$ & $0.45^{+0.06~+0.12~+0.18}_{-0.06~-0.12~-0.18}$ \\ 
				
				R2 &  & $69.29^{+0.37~+0.72~+1.10}_{-0.37~-0.73~-1.10}$ & $~0.07^{+0.08~+0.15~+0.23}_{-0.08~-0.16~-0.24}$  & $0.40^{+0.03~+0.06~+0.09}_{-0.03~-0.06~-0.09}$\\ 
				
				\hline	
				
				N1 &  & $68.91^{+0.56~+1.13~+1.72}_{-0.56~-1.10~-1.65}$ & $-0.15^{+0.10~+0.21~+0.32}_{-0.10~-0.20~-0.31}$ & $0.50^{+0.05~+0.11~+0.15}_{-0.05~-0.10~-0.16}$\\ 
				
				N2 &  & $68.81^{+0.35~+0.69~+1.06}_{-0.35~-0.68~-1.03}$ & $-0.04^{+0.08~+0.15~+0.22}_{-0.07~-0.14~-0.22}$ & $0.43^{+0.05~+0.09~+0.14}_{-0.05~-0.09~-0.14}$\\ 
				
				P1 & Mat. & $68.86^{+0.17~+0.33~+0.49}_{-0.17~-0.33~-0.49}$ & $~0.07^{+0.06~+0.11~+0.17}_{-0.06~-0.11~-0.17}$ & $0.51^{+0.05~+0.10~+0.15}_{-0.05~-0.10~-0.15}$\\ 
				
				P2 & 9/2 & $69.32^{+0.12~+0.24~+0.36}_{-0.12~-0.24~-0.36}$ & $~0.05^{+0.04~+0.08~+0.12}_{-0.04~-0.08~-0.13}$ & $0.40^{+0.04~+0.08~+0.13}_{-0.04~-0.08~-0.13}$\\ 
				
				R1 &  & $68.63^{+0.55~+1.09~+1.64}_{-0.53~-1.04~-1.57}$ & $-0.05^{+0.11~+0.21~+0.31}_{-0.10~-0.20~-0.30}$ & $0.52^{+0.05~+0.12~+0.17}_{-0.05~-0.11~-0.16}$ \\ 
				
				R2 &  & $69.04^{+0.32~+0.64~+0.96}_{-0.32~-0.63~-0.95}$ & $-0.01^{+0.07~+0.14~+0.21}_{-0.07~-0.14~-0.21}$ & $0.43^{+0.04~+0.08~+0.12}_{-0.04~-0.08~-0.12}$\\ 
				
				\hline	
				
				N1 &  & $69.59^{+0.58~+1.15~+1.77}_{-0.57~-1.13~-1.68}$ & $-0.19^{+0.10~+0.20~+0.31}_{-0.10~-0.20~-0.29}$ & $0.41^{+0.07~+0.13~+0.20}_{-0.07~-0.13~-0.21}$\\ 
				
				N2 &  & $68.91^{+0.33~+0.66~+0.99}_{-0.33~-0.64~-0.97}$ & $-0.06^{+0.07~+0.14~+0.21}_{-0.07~-0.14~-0.21}$ & $0.41^{+0.04~+0.07~+0.11}_{-0.04~-0.07~-0.11}$\\ 
				
				P1 &  & $68.94^{+0.16~+0.32~+0.49}_{-0.16~-0.32~-0.49}$ & $~0.07^{+0.06~+0.11~+0.17}_{-0.06~-0.12~-0.17}$ & $0.45^{+0.07~+0.13~+0.19}_{-0.07~-0.14~-0.19}$\\ 
				
				P2 & Cauchy & $69.35^{+0.12~+0.23~+0.35}_{-0.12~-0.23~-0.35}$ & $~0.05^{+0.04~+0.08~+0.12}_{-0.04~-0.08~-0.13}$ & $0.38^{+0.04~+0.07~+0.11}_{-0.04~-0.07~-0.11}$\\ 
				
				R1 &  & $70.23^{+0.55~+1.09~+1.67}_{-0.54~-1.06~-1.59}$ & $-0.10^{+0.10~+0.20~+0.30}_{-0.10~-0.19~-0.29}$ & $0.38^{+0.07~+0.15~+0.21}_{-0.07~-0.14~-0.20}$\\ 
				
				R2 &  & $69.18^{+0.33~+0.64~+0.96}_{-0.32~-0.63~-0.95}$ & $-0.01^{+0.07~+0.14~+0.21}_{-0.07~-0.14~-0.21}$ & $0.39^{+0.04~+0.07~+0.11}_{-0.04~-0.07~-0.11}$\\ 
				
				\hline
				
				N1 &  & $70.46^{+0.62~+1.22~+1.87}_{-0.62~-1.21~-1.83}$ & $-0.26^{+0.10~+0.20~+0.31}_{-0.10~-0.21~-0.31}$ & $0.37^{+0.07~+0.14~+0.20}_{-0.07~-0.13~-0.20}$ \\ 
				
				N2 &  & $68.95^{+0.34~+0.67~+1.02}_{-0.33~-0.65~-0.99}$ & $-0.08^{+0.08~+0.15~+0.22}_{-0.07~-0.14~-0.22}$ & $0.41^{+0.04~+0.07~+0.11}_{-0.04~-0.08~-0.11}$ \\ 
				
				P1 & Rat. & $68.99^{+0.17~+0.34~+0.52}_{-0.17~-0.34~-0.52}$ & $~0.06^{+0.06~+0.12~+0.18}_{-0.06~-0.12~-0.19}$ & $0.45^{+0.07~+0.13~+0.20}_{-0.07~-0.13~-0.20}$ \\ 
				
				P2 & Quad. & $69.35^{+0.13~+0.25~+0.38}_{-0.13~-0.25~-0.38}$ & $~0.04^{+0.05~+0.09~+0.14}_{-0.05~-0.09~-0.14}$ & $0.38^{+0.04~+0.08~+0.12}_{-0.04~-0.08~-0.12}$ \\ 
				
				R1 &  & $71.03^{+0.57~+1.22~+1.70}_{-0.56~-1.11~-1.68}$ & $-0.19^{+0.10~+0.20~+0.30}_{-0.10~-0.20~-0.30}$ & $0.37^{+0.06~+0.12~+0.17}_{-0.06~-0.12~-0.17}$ \\ 
				
				R2 &  & $69.23^{+0.34~+0.66~+1.00}_{-0.33~-0.64~-0.97}$ & $-0.03^{+0.08~+0.15~+0.22}_{-0.07~-0.14~-0.22}$ & $0.39^{+0.04~+0.07~+0.11}_{-0.04~-0.07~-0.11}$ \\ 
				
				\hline
				
			\end{tabular}
		}
	\end{center}
	\label{HOk_res}
\end{table*}

\section{Thermodynamic consistency of $\Omega_{k0}$ constraints}

In this section, the consistency of constraints obtained on $\Omega_{k0}$ with the second law of thermodynamics is looked at. We assume the universe as a system 
is bounded by a cosmological horizon, and the matter content of the universe is enclosed within a volume defined by a radius not bigger than the horizon \cite{gibbons, 
	jacob, padma}. In cosmology, the apparent horizon  $r_A$ serves as the cosmological horizon, which is given by the equation $g^{\mu \nu} R_{,\mu} R_{,\nu} = 0$, 
where $R = a(t)r$ is the proper radius of the 2-sphere and $r$ is the comoving radius. For the FLRW universe with a spatial curvature index $k$, the apparent horizon 
is thus given by 

\begin{equation} 
r_A = \left({H^2 + \frac{k}{a^2}}\right)^{-\frac{1}{2}}. \label{r_A} 
\end{equation} 

For $k = 0$, the apparent horizon reduces to the Hubble horizon $r_H = \frac{1}{H}$. \\

Now, the entropy of the horizon $S_A$ can be written as \cite{bak},

\begin{equation} \label{S_A}
S_A =  8 \pi^2 r_A^2 = \frac{8 \pi^2}{H^2 + \frac{k}{a^2}}. 
\end{equation}

For the second law to be valid, the entropy $S$ should be non-decreasing with respect to the expansion of the universe. If $S_f$ and $S_A$ stand for the entropy 
of the fluid describing the observable universe, and that of the horizon containing the fluid, respectively, then the total entropy of the system, i.e., $S = S_f 
+ S_A$, should satisfy the relation 
\begin{equation} 
\frac{d S}{da} \equiv \frac{d S_f}{da} + \frac{d S_A}{da} \geq 0. \label{2ndLaw} 
\end{equation}

Recently Ferreira and Pav\'{o}n\cite{pavon} gave a prescription to ascertain the signature of $k$ from the second law of thermodynamics. It is a fair assumption that 
the entropy of the observable universe is dominated by that of the cosmic horizon $S_A$ \cite{f_p_21}. So, the second law can be safely written as \cite{pavon}, 
\begin{equation} 
\frac{d S_A}{d a} \geq 0. \label{2ndLaw_approx} 
\end{equation}

Using equation \eqref{S_A} in \eqref{2ndLaw_approx} one can obtain the following condition, 

\begin{equation}
H \frac{d H}{da} \leq \frac{k}{a^3}. \label{cond1}
\end{equation}

The inequality \eqref{cond1} can be rewritten, with a bit of simple algebraic exercise, as 

\begin{equation} 
1 + q \geq \Omega_{k}. \label{cond2}
\end{equation} 

Here $q$ is the deceleration parameter which gives a dimensionless measure of the cosmic acceleration and is defined as, 

\begin{equation}
q = -\frac{\ddot{a}}{aH^2} = -1 + (1+z)\frac{H'}{H}.
\end{equation}

Testing the thermodynamic validity for the obtained constraints on $\Omega_{k0}$ requires a reconstruction of $q_0$ from the respective combination of data sets. 
Quite a lot of work on a non-parametric reconstruction of the cosmic deceleration parameter $q$ is already there in the literature. Some of them can be found in 
\cite{lin_q, xia_q, purba_q, jesus_q, purba_j, keeley2020}. The list, however, is far from being exhaustive. We use the same datasets, that were used for the 
reconstruction of $\Omega_{k0}$, to find the corresponding values of $q_0$.  A very brief methodology is the following. For a more detailed technical description, 
we refer to \cite{purba_int, purba_q}. The comoving distance $D(z)$, and its derivatives $D'(z)$ and $D''(z)$ are reconstructed w.r.t $z$ for different combinations 
of data sets. The uncertainty in $D(z)$ from the corresponding data set is taken into account. For the CC and $r$BAO data, we convert the $H$-$\sigma_H$ data to 
$E$-$\sigma_E$ data set using Eq. \eqref{h_recon} and \eqref{sigh_recon}. $D'(z)$ is then connected to $D(z)$ and $E(z)$ via Eq. \eqref{D} as, 

\begin{equation}
D'(z) = \frac{\sqrt{1+\Omega_{k0} D^2(z)}}{E(z)}.
\end{equation}

Thus, we take into account the $E$ data points, the uncertainty associated $\sigma_E$ while performing the GP reconstruction. We add two extra points $D(0)=0$ to 
the $D$ data set, and $E(0) = 1$ to the $E$ data before proceeding with the reconstruction. We obtain the reconstructed values of $D(z)$, $D'(z)$ and $D''(z)$ at 
the present epoch, along with their error uncertainties. Now, $q$ can be rewritten as a function of $D(z)$ and its derivatives as, 

\begin{equation}
q(z) = -1 + \frac{\Omega_{k0}DD'^2 - (1+\Omega_{k0}D^2)D''}{D'(1+\Omega_{k0}D^2)}(1+z) .
\end{equation}

Using the reconstructed $D(0)$, $D'(0)$ and $D''(0)$, we obtain the values of $1+q_0$, shown in the third column of Table \ref{HOk_res}. Eq. \eqref{cond2} 
asserts that $\Omega_{k0} \leq 1+q_0$ for the second law to be valid. We see that for all combinations, the second law is satisfied by a generous margin, 
independent of the choice of the kernel.

\section{Reconstruction along with the Perturbation data}

Redshift-space distortions are an effect in observational cosmology where the spatial distribution of galaxies appears distorted when their positions are 
looked at as a function of their redshift, rather than as functions of their distances. This effect occurs due to the peculiar velocities of the galaxies 
causing a Doppler shift in addition to the redshift caused by the cosmological expansion. The growth of large structure can not only probe the background 
evolution of the universe, but also distinguish between GR and different modified  gravity theories \cite{ref39, ref41}. Recently, non-parametric 
constraints on the Hubble parameter $H$ and the matter density parameter $\Omega_{m}$ were obtained using the data from cosmic chronometers, type-Ia 
supernovae, baryon acoustic oscillations and redshift-space distortions, assuming a spatially flat universe \cite{ref_new2022}. In this section, we 
propose a non-parametric method to use the growth rate data measured from RSDs to constrain the spatial curvature. \\

\begin{table*}[t!]
	\caption{{\small The best fit values of $\Omega_{m0}$, $\Omega_{k0}$, $\sigma_{8,0}$ and $\gamma$ along with their 1$\sigma$, $2\sigma$ and 3$\sigma$ 
			uncertainties from different combinations of datasets for four choices of the covariance function using the background and perturbation data.}}
	\begin{center}
		\resizebox{\textwidth}{!}{\renewcommand{\arraystretch}{1.55} \setlength{\tabcolsep}{9.5pt} \centering  
			\begin{tabular}{l c c c c c } 
				\hline 	
				\textbf{Set} & $\kappa(z,\tilde{z})$ & $\Omega_{m0}$ &  $\Omega_{k0}$ & $\sigma_{8,0}$ & $\gamma$\\ 
				\hline \hline
				N3 &  & $0.204^{+0.042~+0.082~+0.126}_{-0.041~-0.079~-0.121}$ & $0.040^{+0.152~+0.285~+0.419}_{-0.161~-0.313~-0.483}$ & $0.952^{+0.074~+0.163~+0.296}_{-0.063~-0.116~-0.171}$ & $0.629^{+0.053~+0.139~+0.311}_{-0.045~-0.094~-0.148}$ \\ 
				
				N4 &   & $0.196^{+0.023~+0.044~+0.068}_{-0.023~-0.045~-0.065}$ & $-0.097^{+0.105~+0.205~+0.299}_{-0.102~-0.202~-0.314}$ & $0.964^{+0.039~+0.082~+0.120}_{-0.034~-0.063~-0.090}$ & $0.619^{+0.020~+0.040~+0.066}_{-0.020~-0.041~-0.079}$\\ 
				
				P3 & Sq.  & $0.199^{+0.042~+0.083~+0.125}_{-0.041~-0.075~-0.099}$ & $0.077^{+0.084~+0.159~+0.212}_{-0.091~-0.178~-0.265}$  & $0.961^{+0.077~+0.159~+0.232}_{-0.065~-0.118~-0.164}$ & $0.626^{+0.049~+0.102~+0.174}_{-0.049~-0.095~-0.132}$ \\ 
				
				P4 & Exp.  & $0.185^{+0.021~+0.041~+0.066}_{-0.021~-0.043~-0.063}$ & $0.007^{+0.065~+0.127~+0.218}_{-0.063~-0.122~-0.186}$ & $0.976^{+0.040~+0.089~+0.141}_{-0.036~-0.067~-0.100}$ &  $0.618^{+0.023~+0.047~+0.103}_{-0.023~-0.050~-0.078}$ \\ 
				
				R3 &   & $0.203^{+0.032~+0.064~+0.096}_{-0.033~-0.064~-0.108}$  & $0.078^{+0.073~+0.144~+0.212}_{-0.077~-0.155~-0.232}$ & $0.885^{+0.063~+0.132~+0.250}_{-0.052~-0.098~-0.144}$ & $0.688^{+0.058~+0.139~+0.278}_{-0.050~-0.098~-0.156}$ \\ 
				
				R4 &   & $0.159^{+0.020~+0.042~+0.077}_{-0.021~-0.041~-0.056}$ & $0.005^{+0.061~+0.115~+0.171}_{-0.060~-0.119~-0.228}$ & $0.963^{+0.045~+0.097~+0.148}_{-0.040~-0.076~-0.117}$ &  $0.625^{+0.026~+0.052~+0.079}_{-0.027~-0.056~-0.081}$\\ 
				
				\hline
				
				N3 &  & $0.227^{+0.041~+0.079~+0.117}_{-0.040~-0.074~-0.105}$ & $-0.110^{+0.155~+0.307~+0.445}_{-0.155~-0.312~-0.498}$ & $0.897^{+0.057~+0.118~+0.183}_{-0.048~-0.089~-0.125}$ & $0.615^{+0.036~+0.072~+0.118}_{-0.034~-0.066~-0.097}$\\ 
				
				N4 &   & $0.227^{+0.027~+0.053~+0.079}_{-0.026~-0.049~-0.074}$ & $-0.026^{+0.104~+0.202~+0.297}_{-0.104~-0.215~-0.322}$ & $0.897^{+0.039~+0.076~+0.125}_{-0.034~-0.064~-0.090}$ & $0.633^{+0.025~+0.052~+0.082}_{-0.026~-0.050~-0.078}$ \\ 
				
				P3 &  Mat.  &  $0.228^{+0.038~+0.077~+0.114}_{-0.038~-0.077~-0.108}$ & $-0.044^{+0.089~+0.175~+0.264}_{-0.090~-0.176~-0.270}$ & $0.903^{+0.055~+0.125~+0.195}_{-0.048~-0.091~-0.127}$ & $0.615^{+0.037~+0.073~+0.114}_{-0.035~-0.073~-0.107}$ \\ 
				
				P4 & $9/2$  & $0.221^{+0.025~+0.049~+0.076}_{-0.024~-0.047~-0.066}$ & $0.018^{+0.067~+0.128~+0.185}_{-0.067~-0.131~-0.196}$ & $0.902^{+0.038~+0.077~+0.115}_{-0.035~-0.065~-0.095}$ & $0.632^{+0.027~+0.054~+0.083}_{-0.026~-0.051~-0.072}$ \\ 
				
				R3 &   & $0.220^{+0.032~+0.064~+0.100}_{-0.032~-0.062~-0.093}$ & $-0.031^{+0.083~+0.159~+0.235}_{-0.081~-0.159~-0.242}$ & $0.853^{+0.048~+0.102~+0.171}_{-0.042~-0.079~-0.115}$ & $0.651^{+0.039~+0.080~+0.125}_{-0.036~-0.070~-0.107}$ \\ 
				
				R4 &   & $0.178^{+0.024~+0.049~+0.076}_{-0.023~-0.045~-0.065}$ & $0.034^{+0.065~+0.124~+0.177}_{-0.063~-0.123~-0.188}$ & $0.907^{+0.046~+0.096~+0.147}_{-0.040~-0.076~-0.110}$ & $0.629^{+0.031~+0.061~+0.093}_{-0.030~-0.061~-0.094}$ \\ 
				
				\hline
				
				N3 &  & $0.245^{+0.040~+0.078~+0.120}_{-0.038~-0.073~-0.115}$ & $-0.227^{+0.146~+0.291~+0.426}_{-0.164~-0.318~-0.494}$ & $0.855^{+0.047~+0.099~+0.169}_{-0.041~-0.075~-0.107}$ &  $0.608^{+0.029~+0.059~+0.089}_{-0.029~-0.056~-0.091}$ \\ 
				
				N4 &   & $0.278^{+0.027~+0.053~+0.080}_{-0.026~-0.050~-0.076}$ & $-0.015^{+0.104~+0.198~+0.295}_{-0.108~-0.208~-0.314}$ & $0.820^{+0.029~+0.061~+0.097}_{-0.028~-0.052~-0.076}$ & 
				$0.663^{+0.028~+0.060~+0.099}_{-0.026~-0.048~-0.071}$ \\ 
				
				P3 &   & $0.246^{+0.039~+0.076~+0.115}_{-0.039~-0.082~-0.132}$ & $-0.137^{+0.092~+0.183~+0.285}_{-0.091~-0.180~-0.273}$ & $0.868^{+0.049~+0.114~+0.229}_{-0.042~-0.077~-0.110}$ & $0.602^{+0.031~+0.062~+0.098}_{-0.031~-0.065~-0.127}$ \\ 
				
				P4 & Cauchy & $0.275^{+0.024~+0.048~+0.073}_{-0.024~-0.047~-0.071}$ & $0.017^{+0.063~+0.124~+0.183}_{-0.067~-0.130~-0.196}$ & $0.821^{+0.029~+0.060~+0.095}_{-0.027~-0.051~-0.075}$ & $0.665^{+0.027~+0.055~+0.086}_{-0.026~-0.050~-0.074}$ \\ 
				
				R3 &   & $0.238^{+0.032~+0.063~+0.092}_{-0.032~-0.063~-0.092}$ & $-0.123^{+0.081~+0.159~+0.246}_{-0.080~-0.161~-0.243}$ & $0.823^{+0.041~+0.087~+0.136}_{-0.036~-0.067~-0.096}$ & $0.633^{+0.031~+0.061~+0.097}_{-0.030~-0.059~-0.090}$ \\ 
				
				R4 &   & $0.236^{+0.024~+0.048~+0.072}_{-0.024~-0.045~-0.072}$ & $0.022^{+0.059~+0.115~+0.169}_{-0.061~-0.119~-0.179}$ & $0.807^{+0.033~+0.067~+0.112}_{-0.030~-0.057~-0.083}$ & $0.679^{+0.031~+0.063~+0.102}_{-0.030~-0.058~-0.092}$ \\ 
				
				\hline				
				
				N3 &  & $0.271^{+0.039~+0.079~+0.120}_{-0.040~-0.079~-0.118}$ & $-0.305^{+0.155~+0.298~+0.453}_{-0.163~-0.333~-0.509}$ & $0.799^{+0.039~+0.083~+0.133}_{-0.034~-0.064~-0.094}$ & $0.599^{+0.025~+0.051~+0.096}_{-0.024~-0.049~-0.075}$\\ 
				
				N4 &   &  $0.426^{+0.045~+0.090~+0.135}_{-0.045~-0.090~-0.169}$ & $0.049^{+0.117~+0.236~+0.375}_{-0.121~-0.235~-0.351}$ & $0.657^{+0.027~+0.057~+0.095}_{-0.024~-0.045~-0.066}$ & 
				$0.819^{+0.050~+0.118~+0.239}_{-0.042~-0.081~-0.119}$ \\ 
				
				P3 & Rat.  & $0.280^{+0.040~+0.078~+0.119}_{-0.040~-0.078~-0.124}$ & $-0.186^{+0.096~+0.184~+0.269}_{-0.098~-0.191~-0.292}$ & $0.811^{+0.039~+0.082~+0.141}_{-0.035~-0.063~-0.092}$ &  $0.594^{+0.026~+0.053~+0.082}_{-0.026~-0.052~-0.082}$\\ 
				
				P4 & Quad.  & $0.367^{+0.024~+0.048~+0.073}_{-0.024~-0.047~-0.072}$ & $0.110^{+0.070~+0.140~+0.208}_{-0.070~-0.140~-0.218}$ & $0.703^{+0.024~+0.048~+0.074}_{-0.022~-0.043~-0.063}$ & 
				$0.766^{+0.045~+0.101~+0.167}_{-0.038~-0.071~-0.103}$\\ 
				
				R3 &   &  $0.265^{+0.032~+0.062~+0.097}_{-0.032~-0.063~-0.092}$ & $-0.162^{+0.085~+0.165~+0.250}_{-0.090~-0.179~-0.267}$ & $0.777^{+0.033~+0.069~+0.107}_{-0.030~-0.055~-0.083}$ &  $0.618^{+0.025~+0.050~+0.080}_{-0.025~-0.049~-0.073}$\\ 
				
				R4 &   & $0.331^{+0.024~+0.047~+0.070}_{-0.024~-0.047~-0.069}$ & $0.082^{+0.062~+0.124~+0.182}_{-0.062~-0.124~-0.185}$ & $0.672^{+0.023~+0.047~+0.075}_{-0.022~-0.042~-0.060}$ & 
				$0.807^{+0.050~+0.109~+0.170}_{-0.041~-0.076~-0.110}$ \\ 
				\hline
			\end{tabular}
		}
	\end{center}
	\label{Ok_rsd_tab}
\end{table*}

In a background universe filled with matter and dark energy, the evolution of matter density contrast is given by, 

\begin{equation} \label{delta_def}
\delta = \frac{\delta \rho_m}{\rho_m}.
\end{equation} 

In the linearized approximation, $\delta$ obeys the following second order differential equation for its evolution, 

\begin{equation} \label{perturb_eqn}
\ddot{\delta}+ 2H \dot{\delta} - 4\pi G_{\mbox{\tiny eff}} \rho_m \delta = 0,
\end{equation} 

where $\rho_m$ is the background matter density, $\delta \rho_m$ represents its first-order perturbation, and the `dot' denotes derivative
with respect to cosmic time $t$. Note that $G_{\mbox{\tiny eff}}$ is the effective gravitational constant. For Einstein's GR, $G_{\mbox{\tiny eff}}$ 
reduces to the Newton's gravitational constant $G$. Considering the growth factor $f(a) = \frac{d \ln \delta}{d \ln a}$, Gong, Ishak and 
Wang\cite{gong2009} provided an approximate solution to equation \eqref{perturb_eqn} as,

\begin{equation} \label{f_rsd}
f(z) = \Omega_{m}^\gamma + \left(\gamma - \frac{4}{7}\right)\Omega_{k} .
\end{equation} 

Here, $\Omega_{m} = \frac{\Omega_{m0} (1+z)^3}{E^2(z)}$ is the matter density parameter, $\Omega_{k} = \frac{\Omega_{k0} (1+z)^2}{E^2(z)}$ is the 
curvature density parameter and $E(z)= \frac{H(z)}{H_0}$. The growth index $\gamma$ depends on the model. For the $\Lambda$CDM model, 
$f(z) \simeq \Omega_m^{\gamma}$, and $\gamma = 6/11$ is a solution to Eq. \eqref{perturb_eqn} where the terms $\mathcal{O}(1 - \Omega_{m})^2$ are 
neglected \cite{ref35}. For dark energy models with slowly varying equation of state $\gamma \simeq 0.55$ \cite{ref36}. For modified gravity models, 
different values have been predicted in literature, such as $\gamma \simeq 0.68$ for Dvali-Gabadadze-Porrati (DGP) model \cite{ref42, ref43}. The 
RSD data measure the quantity $f \sigma_8$, defined by,

\begin{eqnarray} \label{fs8_theo}
f \sigma_8 (z) &=& f (z)~\sigma_{8,0}~ \frac{\delta(z)}{\delta_0},  \nonumber \\ 
&=& \sigma_{8,0}~f (z)\exp \left\lbrace \int_{0}^{z} - \frac{f(z')}{1+z'} dz'\right\rbrace ,
\end{eqnarray} 

where $\sigma_{8}$ is the linear theory root-mean-square mass fluctuation within a sphere of radius $8h^{-1}$ Mpc \cite{ref21,ref22,ref23,ref24,ref25}, 
$h$ being the dimensionless Hubble parameter at the present epoch. \\

On substituting equation \eqref{f_rsd} in \eqref{fs8_theo} we get,

\begin{widetext}
\begin{equation} \label{fs8_final}
f \sigma_8 (z) = \sigma_{8,0}\left[\Omega_{m}^\gamma + \left(\gamma - \frac{4}{7}\right)\Omega_{k}\right] \exp \left\lbrace \int_{0}^{z} - \frac{\left[\Omega_{m}^\gamma + \left(\gamma - \frac{4}{7}\right)\Omega_{k}\right]}{1+z'} dz'\right\rbrace.
\end{equation} 
\end{widetext}

\begin{figure*}[t!]
	\begin{center}
		\includegraphics[angle=0,width=0.475\textwidth]{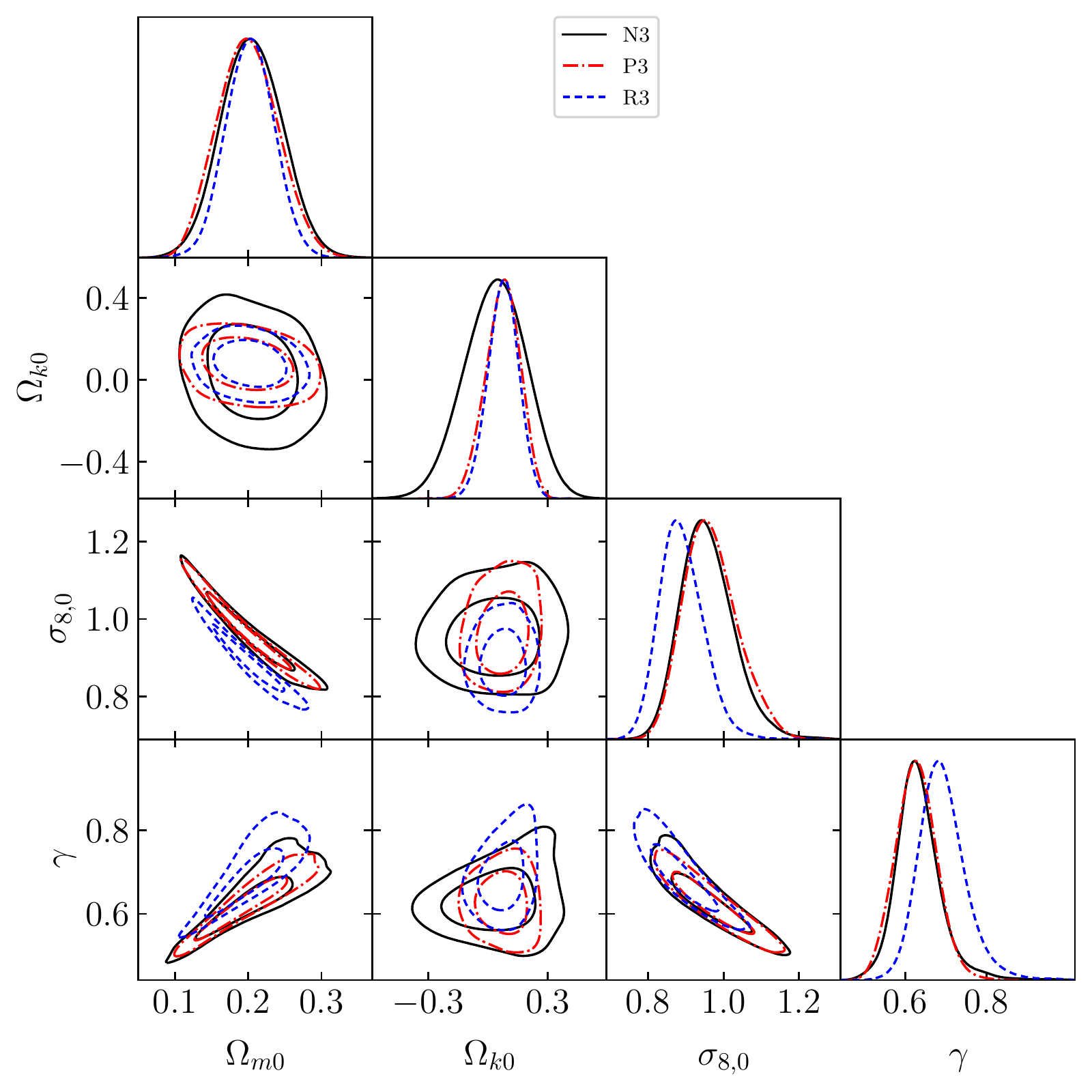}
		\includegraphics[angle=0,width=0.475\textwidth]{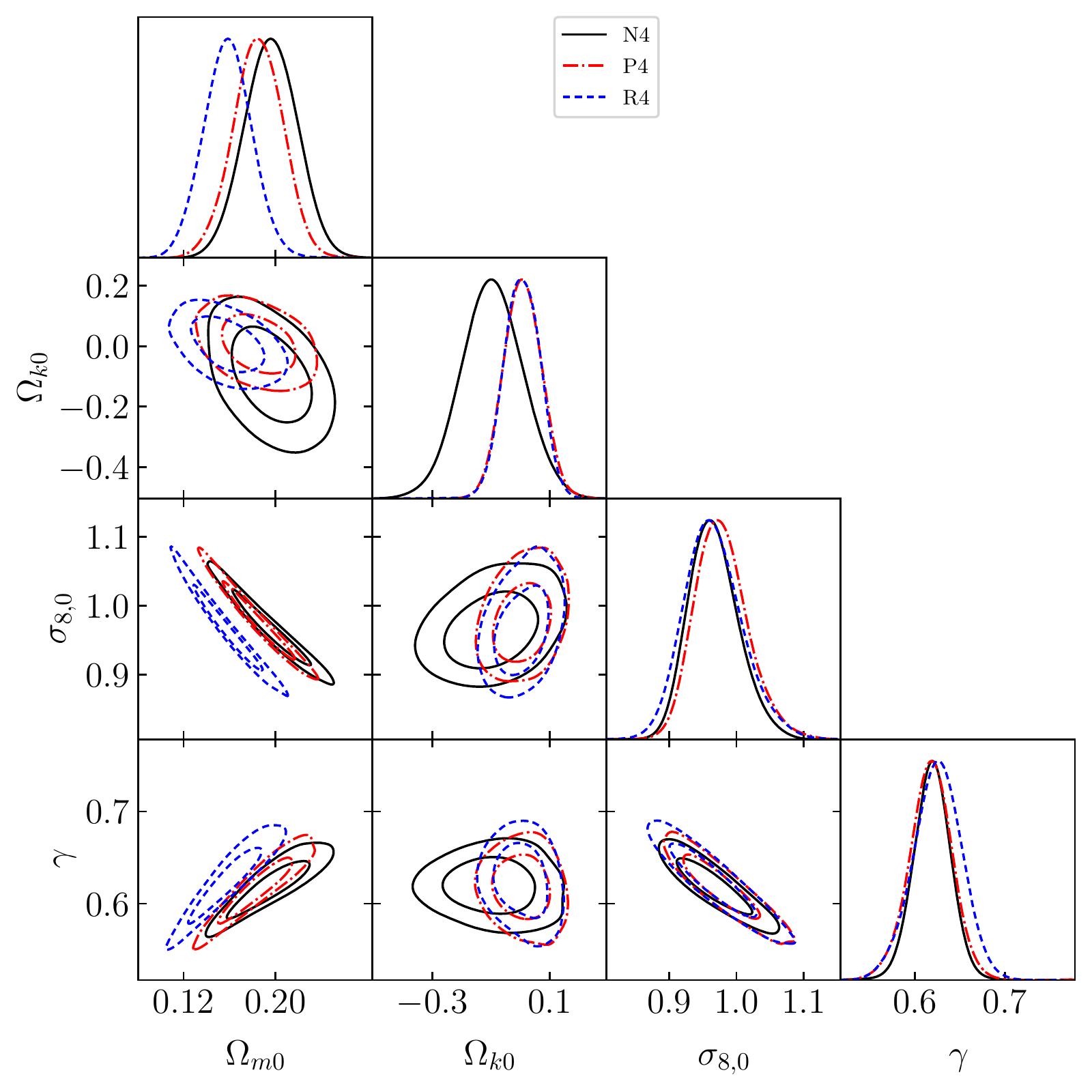}	
	\end{center}
	\caption{{\small Contour plots and the marginalized likelihood of $H_0$ and $\Omega_{k0}$ considering the squared exponential covariance for Set III 
			(left) and Set IV (right). The solid lines represent the results for N3 and N4 data-set combination, dash-dot lines corresponds to the P3 and 
			P4 data-set combination, and the dashed lines  represent the results for R3 and R4 data-set combinations. The associated 1$\sigma$, 2$\sigma$ 
			confidence contours are shown along with the respective marginalized likelihood functions.}}
	\label{Ok_rsd_sqexp}
\end{figure*}

\begin{figure*}[t!]
	\begin{center}
		\includegraphics[angle=0,width=0.475\textwidth]{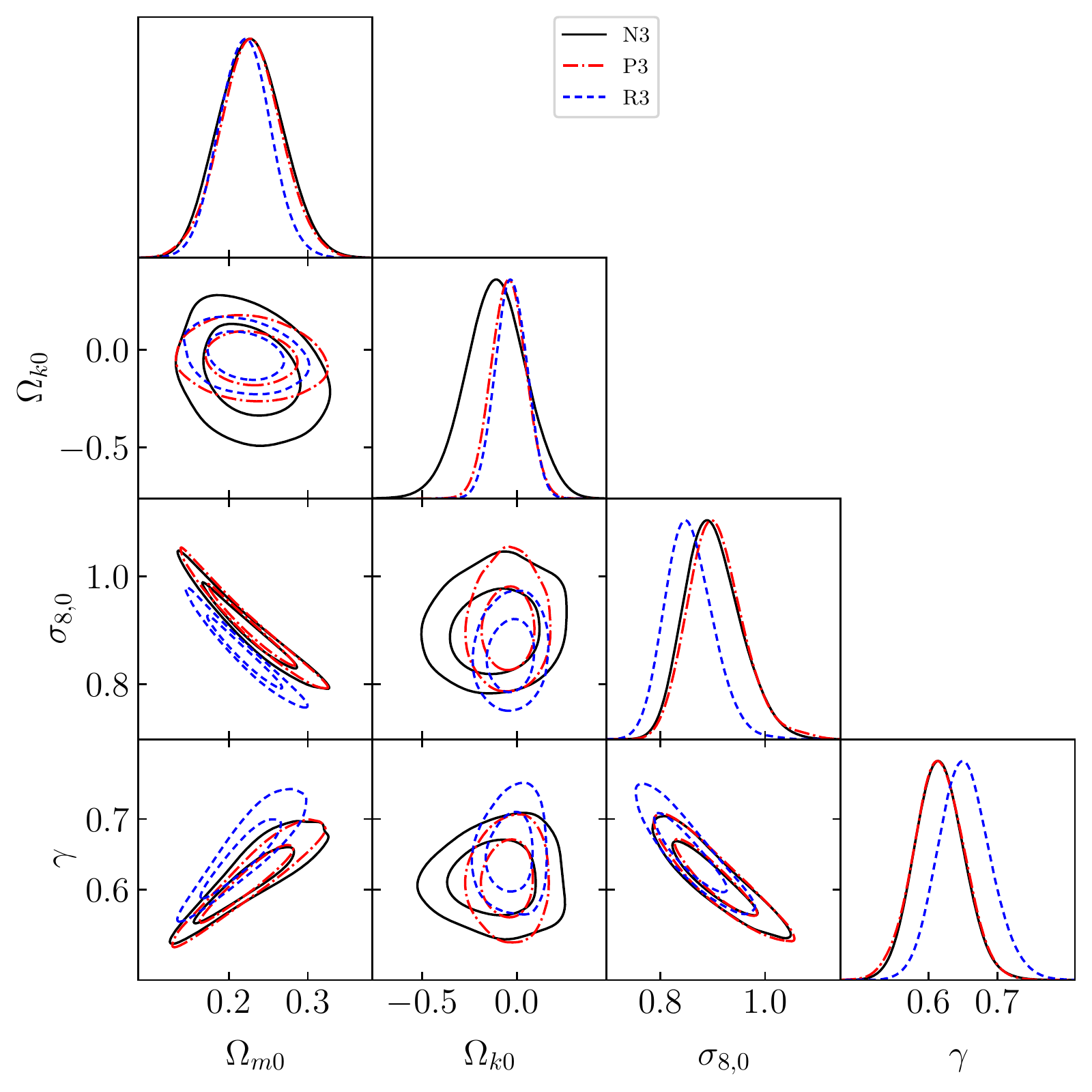}
		\includegraphics[angle=0,width=0.475\textwidth]{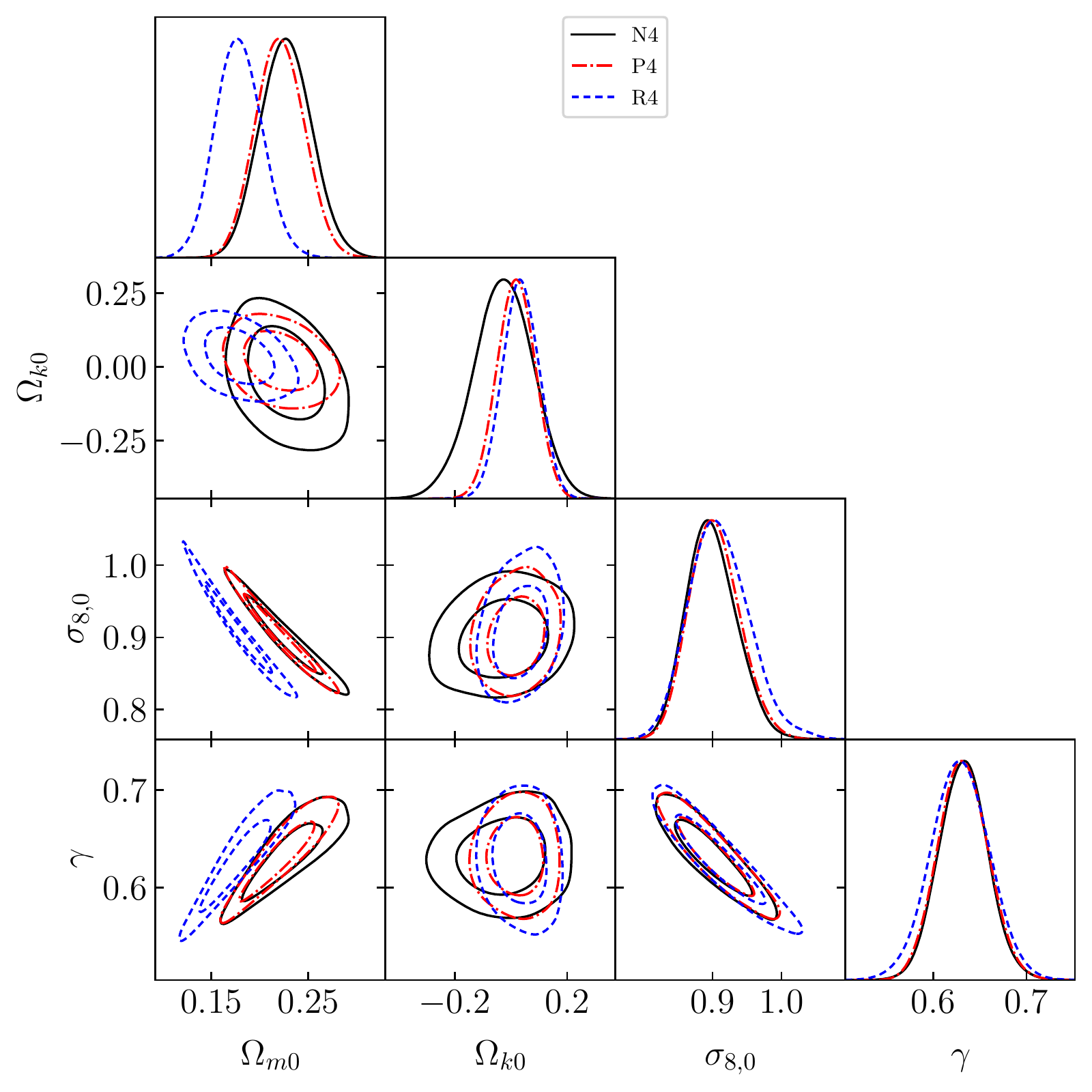}	
	\end{center}
	\caption{{\small Contour plots and the marginalized likelihood of $H_0$ and $\Omega_{k0}$ considering the Mat\'{e}rn $9/2$ covariance for Set III (left) 
			and Set IV (right). The solid lines represent the results for N3 and N4 data-set combination, dash-dot lines corresponds to the P3 and P4 data-set 
			combination, and the dashed lines  represent the results for R3 and R4 data-set combinations. The associated 1$\sigma$, 2$\sigma$ confidence contours 
			are shown along with the respective marginalized likelihood functions.}}
	\label{Ok_rsd_mat92}
\end{figure*}

\begin{figure*}[t!]
	\begin{center}
		\includegraphics[angle=0,width=0.475\textwidth]{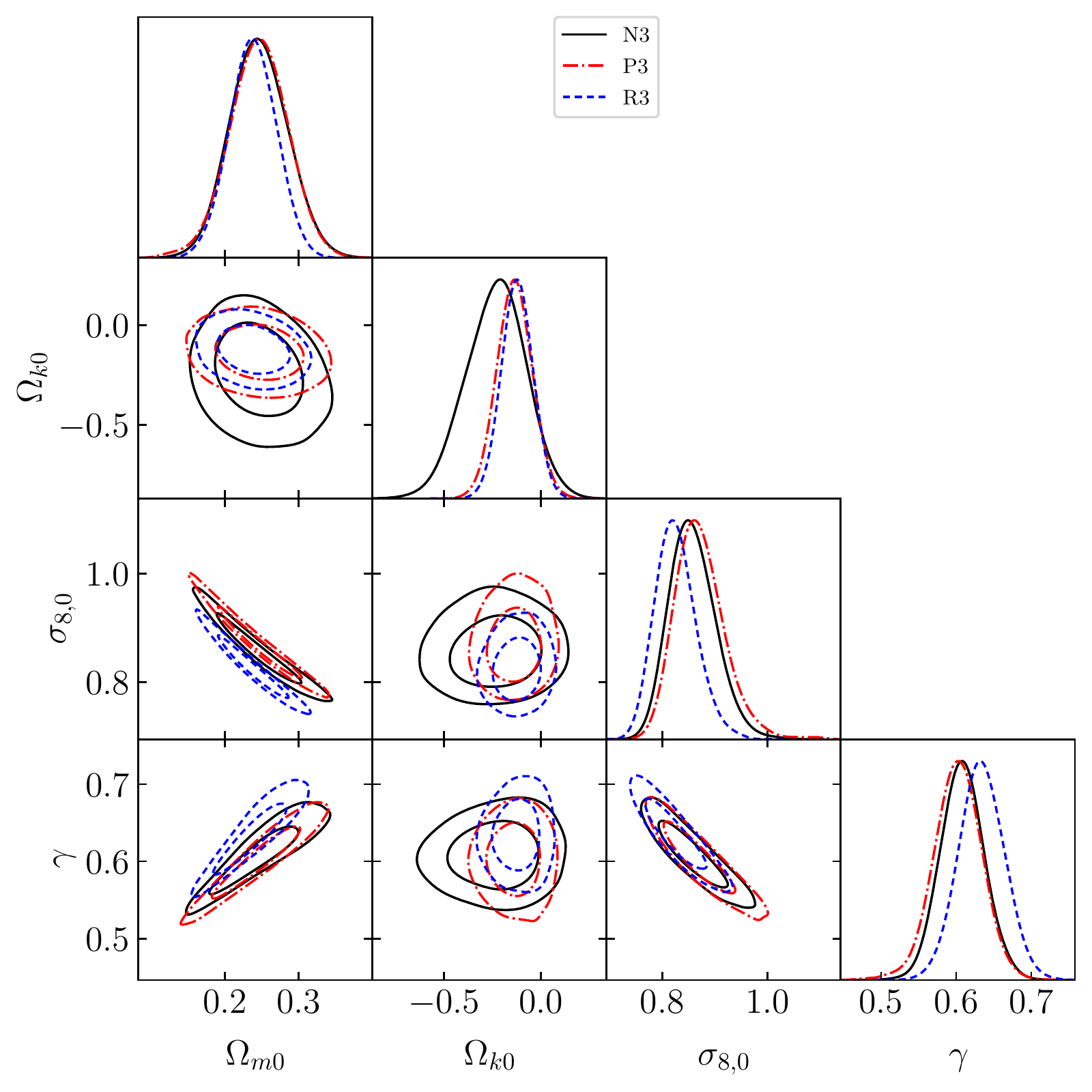}
		\includegraphics[angle=0,width=0.475\textwidth]{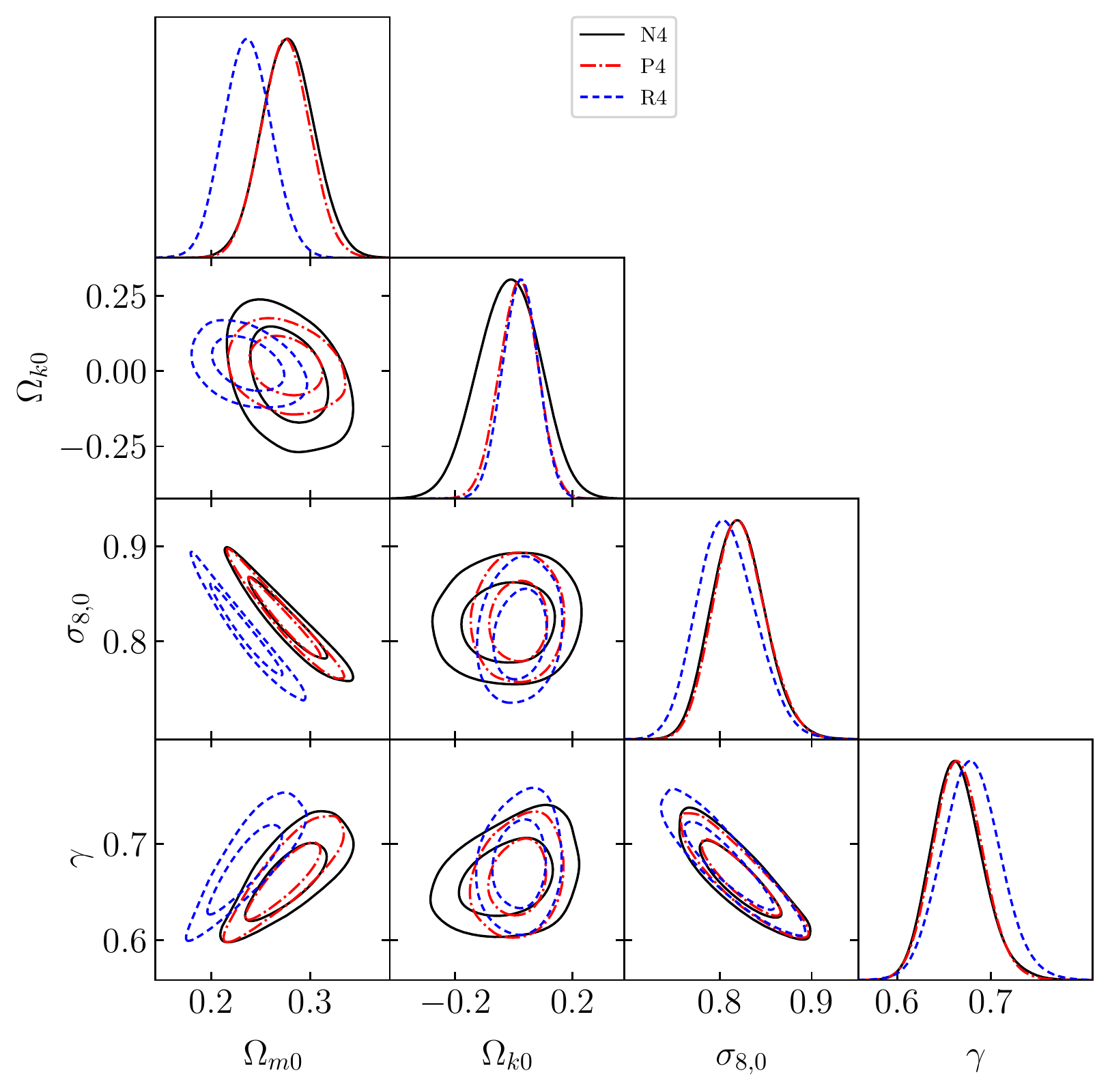}	
	\end{center}
	\caption{{\small Contour plots and the marginalized likelihood of $H_0$ and $\Omega_{k0}$ considering the Cauchy covariance for Set III (left) 
			and Set IV (right). The solid lines represent the results for N3 and N4 data-set combination, dash-dot lines corresponds to the P3 and P4 
			data-set combination, and the dashed lines  represent the results for R3 and R4 data-set combinations. The associated 1$\sigma$, 2$\sigma$ 
			confidence contours are shown along with the respective marginalized likelihood functions.}}
	\label{Ok_rsd_cauchy}
\end{figure*}

\begin{figure*}[t!]
	\begin{center}
		\includegraphics[angle=0,width=0.475\textwidth]{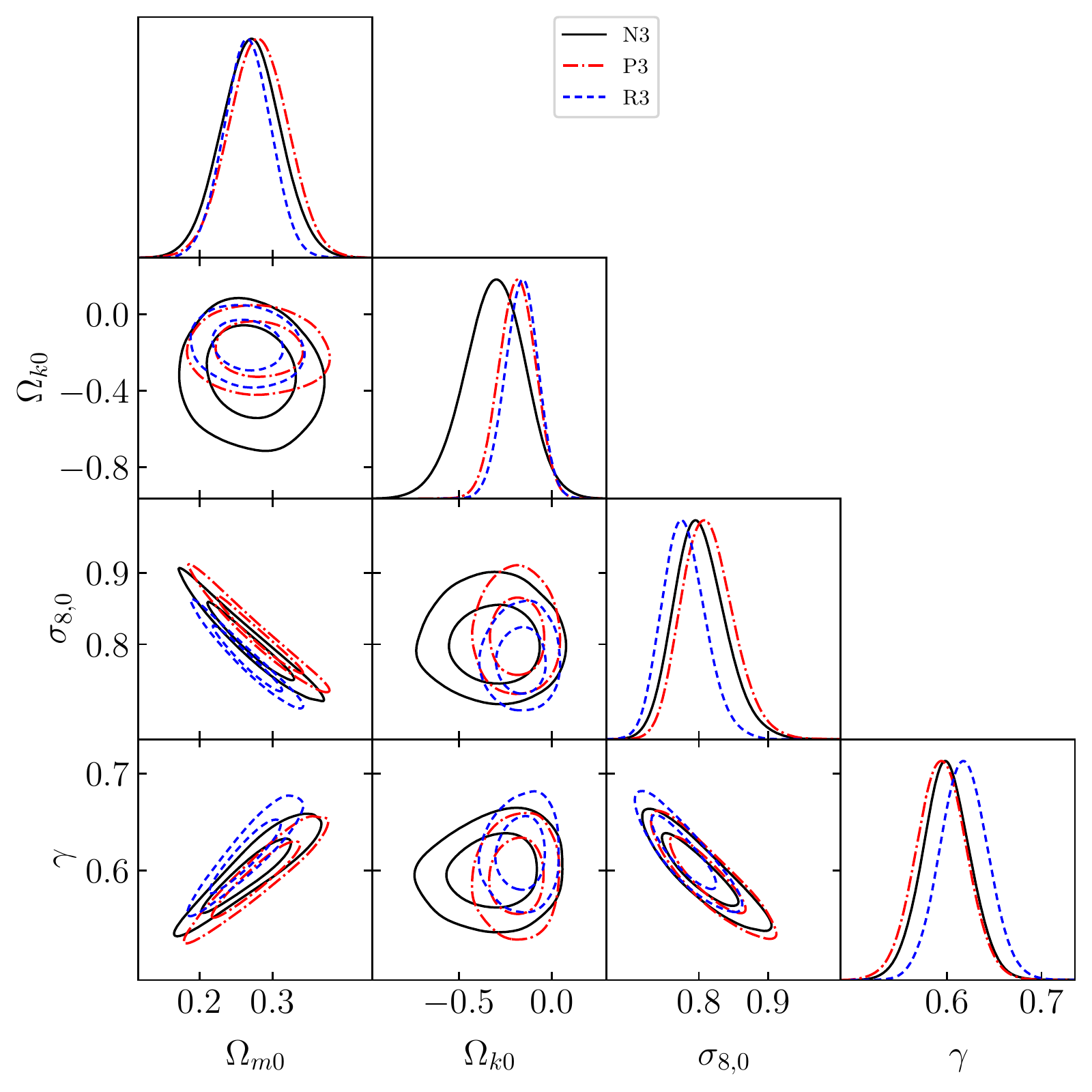}
		\includegraphics[angle=0,width=0.475\textwidth]{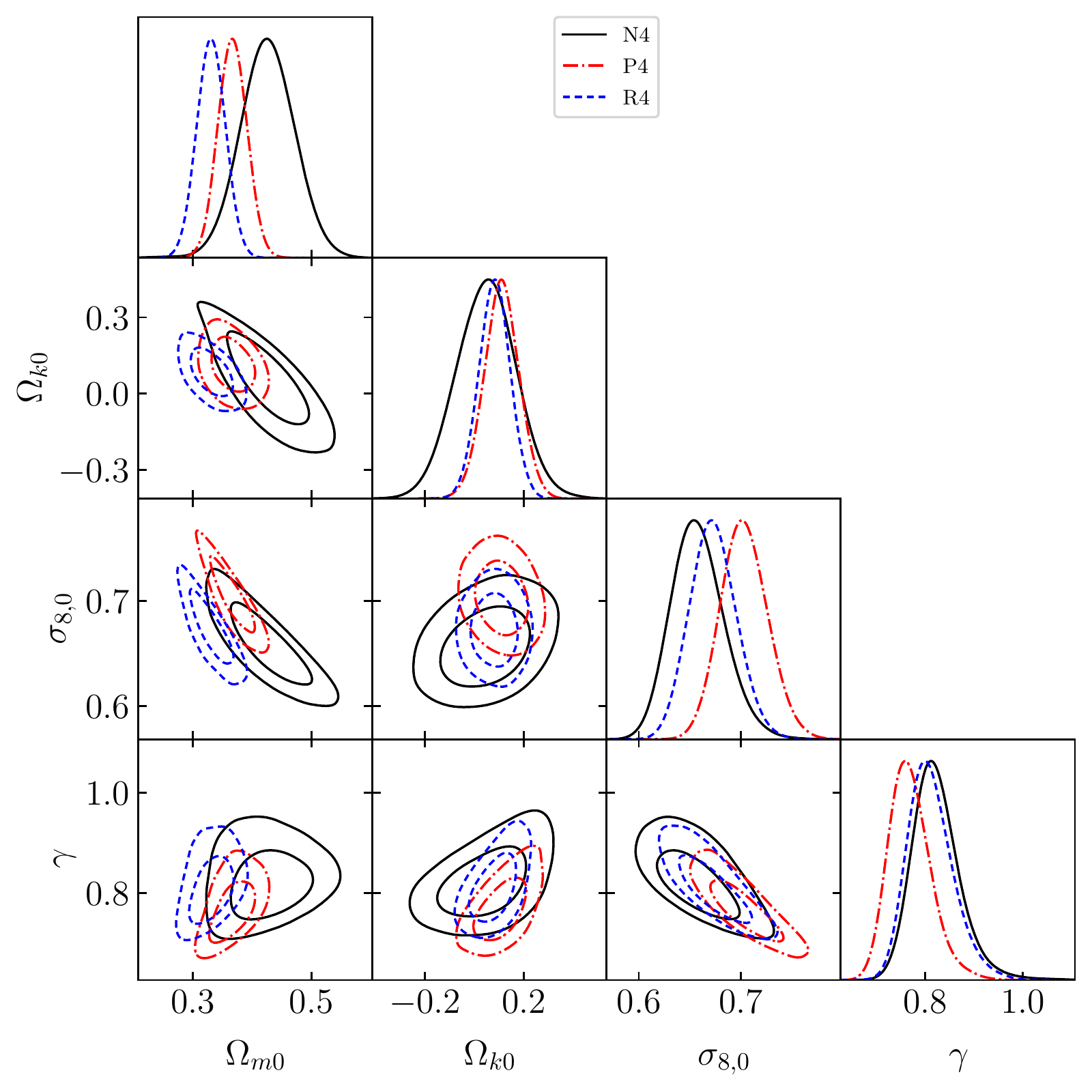}	
	\end{center}
	\caption{{\small Contour plots and the marginalized likelihood of $H_0$ and $\Omega_{k0}$ considering the rational quadratic covariance for Set III 
			(left) and Set IV (right). The solid lines represent the results for N3 and N4 data-set combination, dash-dot lines corresponds to the P3 and P4 
			data-set combination, and the dashed lines  represent the results for R3 and R4 data-set combinations. The associated 1$\sigma$, 2$\sigma$ 
			confidence contours are shown along with the respective marginalized likelihood functions.}}
	\label{Ok_rsd_quad}
\end{figure*}

We proceed with the integration of Eq. \eqref{fs8_final} numerically using the composite trapezoidal rule as in equation \eqref{trap}. The reconstructed $E(z)$ 
function from CC and CC+$r$BAO data are considered. For the Pantheon data, we make use of equations \eqref{D_recon} and \eqref{sigD_recon}. \\

Here we consider the following combination of data sets,
\begin{itemize}
	\item Set III
	\begin{enumerate}
		\item N3 - CC+SN+RSD 
		\item P3 - CC+SN+RSD+P18
		\item R3 - CC+SN+RSD+R19
	\end{enumerate}	
	\item Set IV
	\begin{enumerate}
		\item N4 - CC+$r$BAO+SN+RSD 
		\item P4 - CC+$r$BAO+SN+RSD+P18
		\item R4 - CC+$r$BAO+SN+RSD+R19
	\end{enumerate}	
\end{itemize}

We use the GP method to reconstruct the function $f\sigma_{8}(z)$ from RSD data. Finally, we constrain the cosmological parameters $\Omega_{m0}$, 
$\Omega_{k0}$, $\sigma_{8,0}$ and $\gamma$ utilizing the $\chi^2$ minimization technique. The uncertainties associated are estimated via a Markov 
Chain Monte Carlo analysis. The best fit results along with their respective 1$\sigma$, 2$\sigma$ and 3$\sigma$ uncertainties is given in Table 
\ref{Ok_rsd_tab}. Plots for the marginalized posteriors with 1$\sigma$ and 2$\sigma$ confidence contours using the Set III and Set IV data 
combinations are shown in Figures \ref{Ok_rsd_sqexp}, \ref{Ok_rsd_mat92}, \ref{Ok_rsd_cauchy} and \ref{Ok_rsd_quad}, for the squared exponential, 
Mat\'{e}rn $9/2$, Cauchy and rational quadratic covariance respectively.\\

The marginalized $\Omega_{k0}$ constraints for the N3 combination is consistent with spatial flatness within 1$\sigma$ CL for the squared exponential 
and Mat\'{e}rn 9/2 covariance, within 2$\sigma$ for the Cauchy covariance and within 3$\sigma$ for the rational quadratic covariance. For the N4 
combination, reconstructed $\Omega_{k0}$ lies with 1$\sigma$ for all four kernel choices. Considering the P18 and R19 $H_0$ prior, it is seen that 
the squared exponential kernel includes $\Omega_{k0}=0$ for P3, P4, R4 combinations within 1$\sigma$ and for the R3 combination within 2$\sigma$. The 
Mat\'{e}rn $9/2$ kernel includes $\Omega_{k0}=0$ for all P3, P4, R3, R4 combinations within 1$\sigma$. The Cauchy kernel includes $\Omega_{k0}=0$ for 
the P3, R3 combination in 2$\sigma$, and for the P4, R4 combination in 1$\sigma$ CL. Lastly, utilizing the rational quadratic kernel, $\Omega_{k0}=0$ 
is included in 3$\sigma$ for the P3 and R3 combination, whereas in 2$\sigma$ for the P4, R4 combination. Inclusion of $r$BAO data leads to tighter 
constraints on $\Omega_{k0}$, and the best-fit values are seen to favour a spatially open universe (see Table \ref{Ok_rsd_tab}). \\

The reconstructed values of $\gamma$ show that the $\Lambda$CDM model is mostly included in 2$\sigma$ and always in 3$\sigma$, except for the rational 
quadratic kernel. From Table \ref{Ok_rsd_tab}, it can been seen that for the N4, P4 and R4 combinations, the $\Lambda$CDM model in not included in 
3$\sigma$ considering the rational quadratic kernel, and marginally included in 3$\sigma$ while using the Cauchy covariance.

\section{Discussion}

In the present work, constraints on the cosmic curvature density parameter $\Omega_{k0}$ have been obtained from different cosmological probes with the 
help of a non-parametric reconstruction. The Cosmic Chronometer and the radial Baryon Acoustic Oscillation measurements of the Hubble parameter, the 
recent supernova compilation of the corrected Pantheon sample, along with measurement of the Redshift Space Distortions which measure the growth of large 
structure are utilized for the purpose. The widely used Gaussian Process and the Markov Chain Monte Carlo method have been employed in this work. The 
analysis has been performed for four choices of the covariance function, namely the squared exponential, Mat\'{e}rn $9/2$, Cauchy and rational quadratic 
kernel. The choice of covariance function involves some discretion and thus a bit subjective. The use of various choices of covariance makes the present 
investigation quite exhaustive in that respect. \\

The reconstructed $\Omega_{k0}$ obtained by combining the CC and Pantheon data are consistent with spatial flatness within 1$\sigma$ confidence level for 
the squared exponential covariance function, within 2$\sigma$ CL level for the Mat\'{e}rn 9/2 and Cauchy covariance function, and within 3$\sigma$ CL for 
the rational quadratic covariance. Including the $r$BAO data to the analysis results in tighter constraints on $\Omega_{k0}$. Combining the CC and Pantheon 
data with the BAO data, it can be seen that $\Omega_{k0}$ is consistent with a spatially flat universe at the 1$\sigma$ CL for the squared exponential, 
Mat\'{e}rn 9/2 and Cauchy covariance, whereas in 2$\sigma$ for the rational quadratic kernel. The best-fit values show an inclination towards a closed 
universe in these cases. This result is obtained without using any given $H_0$ priors. We then introduce the P18 and R19 $H_0$ measurements as priors in 
our analysis and examine their effect on the reconstruction. Plots reveal that the best-fit values of $\Omega_{k0}$ favour a spatially open universe for 
the P18 prior choice, whereas the R19 prior favours a spatially closed universe, except for the squared exponential kernel which favours a spatially open 
universe for both the P18 and R19 priors. However, a spatially flat universe is mostly included at 2$\sigma$ CL for both cases (see Table \ref{HOk_res}). 
\\

Consistency with thermodynamic requirements imposed by the generalized second law of thermodynamics for the reconstructed constraints on $\Omega_{k0}$ 
from the background data combinations are checked quite exhaustively. This has been done with the help of the inequality very recently given by Ferreira 
and Pav\'{o}n\cite{pavon} (see also Ref. \cite{pavon2}). It is quite encouraging to see that the constraints obtained are quite consistent with the 
thermodynamic requirements, independent of the choice of the kernel for all possible combinations of data sets (see Table \ref{HOk_res}).\\

In addition to the background data, we also utilize the RSD data to determine $\Omega_{k0}$ using two combination of datasets, CC+Pantheon+RSD and 
CC+$r$BAO+Pantheon+RSD respectively. This inclusion does not help in providing tighter constraints on $\Omega_{k0}$, but is essential as the spatial 
curvature and the formation of large scale structure should be compatible. We also include the R18 and P18 $H_0$ priors and see their effect on the 
reconstruction. The results obtained are consistent with spatial flatness mostly within 2$\sigma$ and always within 3$\sigma$ in the domain of the 
reconstruction, $0<z<2$ (see Table \ref{Ok_rsd_tab}). \\

The GP method has previously been used for constraining $\Omega_{k0}$ from observations. Li \textit{et al.}\cite{li2016} constrained the spatial curvature 
to be $\Omega_{k0} = -0.045^{+0.172}_{-0.172}$ with 22 $H(z)$ and Union 2.1 SN-Ia data, and $\Omega_{k0} = -0.140^{+0.161}_{-0.158}$ considering the JLA 
SN-Ia data, which are in good agreement with a spatially flat universe. Wei \& Wu\cite{wei2017} extended this analysis using different $H_0$ priors and 
showed that the local and global $H_0$ measurements can affect the constraints on $\Omega_{k0}$. Wang \textit{et al.}\cite{wang2017} showed that a spatially 
flat and transparent universe is preferred by observations. The results indicated a strong degeneracy between the curvature parameter and cosmic opacity. 
From 100 simulated GWs signals, Liao\cite{liao2019} found the results favoured a spatially flat universe with $0.057$ uncertainty at 1$\sigma$, which was 
reduced to $0.027$ for 1000 GWs signals. On combining with the SN-Ia data from DES, the uncertainty was further constrained to  $0.027$ and $0.018$ respectively. 
The analysis by Wei \& Melia\cite{wei2019} suggests that a mildly closed universe ($\Omega_{k0}=-0.918 \pm 0.429$) is preferred at the 1$\sigma$ level using 
quasars and CC data. Recently, Wang, Ma \& Xia\cite{wang2020} found a spatially open universe is favoured at 1$\sigma$ CL using 31 CC-H(z) measurements and 
simulated data form GWs, based on the ANN method. Another non-parametric reconstruction of $\Omega_{k0}$ utilizing different approaches like the 
principal component analysis, genetic algorithms, binning with direct error propagation and the Pad\'e approximation, was carried out by Sapone, Majerotto 
and Nesseris\cite{ref_new2014}. Their results were in good agreement with $\Omega_{k}=0$ at the 1$\sigma$ CL.\\

Our work is similar to the recent works by Yang \& Gong\cite{yang2021} and Dhawan, Alsing \& Vagonzzi\cite{dhawan2021}, but there are quite a few differences 
to list. Yang and Gong\cite{yang2021}, Dhawan, Alsing and Vagnozzi\cite{dhawan2021} have used the Pantheon compilation by Scolnic \textit{et al.}\cite{pan1} 
in their analysis. However, in this work we have utilized the very recent redshift corrected version of Pantheon compilation by Steinhardt, Sneppen and 
Sen\cite{pan_correct}. Yang and Gong\cite{yang2021} reconstructed the quantity $\Omega_{k0} h^2$ so that the discrepancy in the present value of Hubble 
parameter $H_0$ is avoided. Dhawan, Alsing and Vagnozzi\cite{dhawan2021} obtained constraints on $\Omega_{k0}$ independent of the absolute calibration of 
either the SN-Ia or CC measurements. In this particular work, we have obtained constraints on both $\Omega_{k0}$ and $H_0$ form the combined CC+Pantheon 
data, thereby capturing the degeneracy or correlation between them. Yang and Gong\cite{yang2021} imposed a zero mean function, which follows the work Seikel 
\textit{et al.}\cite{gp1} and is similar to our work. Dhawan, Alsing and Vagnozzi\cite{dhawan2021}, on the other hand used a mean non-zero constant prior 
equal to 100, following Shafieloo \textit{et al.}\cite{gp2}. Utilizing solely the background data, Yang \& Gong\cite{yang2021} found the case for a spatially 
open universe from the combined CC and Pantheon data at more than 1$\sigma$ CL considering the squared exponential covariance. The present work starts with a 
zero mean prior similar to \cite{yang2021}, but the best-fit value for the combined CC+Pantheon data (N1) using the same squared exponential kernel favours a 
spatially closed universe, and $\Omega_{k0}=0$ is well included within in 1$\sigma$ CL. This result is similar in nature to that given by Dhawan, Alsing and 
Vagnozzi\cite{dhawan2021} where the obtained constraints on $\Omega_{k0} = -0.03 \pm 0.26$ are consistent with spatial flatness at the $\mathcal{O}(10^{-1})$ 
level. The qualitative difference of the present result with that obtained in \cite{yang2021} can stem from the fact that we have used the redshift corrected 
version of the Pantheon compilation\cite{pan_correct}.\\

Our conclusion is that although there is indeed a scope of revisiting the notion of a spatially flat universe, but the present state of affairs is still quite 
consistent with $k=0$. Observations from future surveys, as well as more data on high redshift observations of CC, SN, BAO and other observables should be able 
to provide tighter constraints on $\Omega_{k0}$.


\label{lastpage}

\end{document}